\documentclass[prd,10pt,aps,twocolumn,floatfix]{revtex4-2}
\usepackage{graphicx,epstopdf}
\usepackage[caption=false]{subfig}
\usepackage{appendix}
\usepackage[T1]{fontenc}
\usepackage{lmodern}
\usepackage[dvipsnames,x11names]{xcolor}
\usepackage[psdextra,colorlinks=true,linkcolor=Magenta!80!black,
citecolor=Magenta!70!black,urlcolor=Magenta!80!black]{hyperref}
\usepackage[sort&compress]{natbib}
\usepackage{morefloats}
\usepackage[pdf]{pstricks}
\usepackage{amsmath}
\usepackage{amssymb}
\usepackage{amsfonts}
\usepackage{rotating}
\usepackage{cancel}
\usepackage{mathtools}
\usepackage{bbm}
\usepackage{dsfont}
\usepackage{bbold}
\usepackage{multirow}
\usepackage{ulem}
\usepackage{physics}
\usepackage{orcidlink}
\usepackage{colortbl}
\usepackage{slashed}
\usepackage{booktabs}
 \usepackage{xcolor}
 \usepackage{colortbl}
 \usepackage{array}
\usepackage{dblfloatfix}

\begin{document}

\title{Doubly-strange hidden-charm pentaquarks from the Fermi statistics of the light-quark cloud}

\author{Halil Mutuk}
\email{hmutuk@omu.edu.tr}           
\affiliation{Department of Physics, Faculty of Sciences, Ondokuz Mayis University, 55139 Samsun, Türkiye}     

\date{\today}

\begin{abstract}
We extend the baryo-charmonium picture of pentaquarks---a color-octet
$c\bar c$ core bonded to a color-octet light-quark cloud---to the doubly-strange
sector $c\bar c ssq$. The mass splittings are set entirely by the light cloud,
so the only new inputs are the strange--strange couplings $J^{ss}$, fixed by a
second application of the chromomagnetic scaling, and an additive strange-mass increment taken from the observed $P_c\!\to\!P_{cs}$ shift. We obtain two
negative-parity triplets, one produced with a kaon and one with an antiproton,
the lowest kaon-associated $\tfrac12^-$ state near $4.60$~GeV and the
antiproton-associated triplet some $120$~MeV below. The robust, distinctive
prediction is that the upper two kaon-associated states form a near-degenerate
doublet, in sharp contrast to the well-separated triplets of the lighter
sectors---a sparse, fixed-spacing pattern that sets the scheme apart from the
molecular and diquark alternatives. The internal splittings follow without adjustment from the measured $P_c$ and $P_{cs}$ spectra; 
the absolute scale relies on the additive strange-mass ansatz, the main assumption of the extrapolation. 
The predicted masses agree with recent molecular
coupled-channel and QCD sum-rule results.
\end{abstract}

\maketitle

% ====================================================================
\section{Introduction}
% ====================================================================
Exotic-hadron spectroscopy has advanced rapidly since the LHCb Collaboration
established a family of hidden-charm pentaquarks. The $P_c(4312)$,
$P_c(4440)$, and $P_c(4457)$ were resolved in
$\Lambda_b^0\!\to\! J/\psi p K^-$~\cite{Aaij:2015tga,Aaij:2019vzc}; a strange state
$P_{cs}(4459)$ followed in $\Xi_b^-\!\to\! J/\psi\Lambda K^-$~\cite{LHCb:2020jpq},
and a second strange state $P_{\psi s}^{\Lambda}(4338)$, with measured
$J^P=\tfrac12^-$, in $B^-\!\to\! J/\psi\Lambda\bar p$~\cite{LHCb:2022ogu}; the
latter has since been corroborated by Belle~\cite{Belle:2025pey}. These states have
been interpreted as loosely bound hadronic molecules, as compact diquark
configurations, and within Born--Oppenheimer schemes~\cite{Maiani:2023nwj}; for
reviews see Refs.~\cite{Esposito:2016noz,Meng:2022ozq}.

A complementary picture was put forward recently in Ref.~\cite{Germani:2024miu}, in
which the pentaquark is a ``baryo-charmonium'': three light quarks in a
color-octet $(qqq)_8$ cloud orbiting the mean color field of a color-octet
$(c\bar c)_8$ pair. Its defining feature is that the dynamics responsible for
the observed fine structure resides in the light cloud rather than in the
heavy core. Fermi statistics restricts the allowed color-flavor-spin
configurations of the cloud, and a residual exchange interaction among the
light quarks generates the mass splittings. With a small set of light-quark
couplings fixed by the non-strange and singly-strange data, the scheme
reproduces the masses of all six observed states, orders their spins, and
sorts them into two spectroscopic series distinguished by production
mechanism.

The motivation for this picture is twofold. First, it isolates the mechanism
behind the observed fine structure. In an ordinary octet baryon the three
light quarks are forced into a fully color-antisymmetric state; the three
pairwise couplings are then equal, the spin-exchange splitting vanishes
identically, and only $J=\tfrac12$ survives. The pentaquark cloud, by
contrast, is a color octet rather than a color singlet and is therefore no
longer fully antisymmetric in color: the pairwise couplings need not coincide,
both $J=\tfrac12$ and $J=\tfrac32$ become accessible, and a nontrivial spin
spectrum is produced by the very same exchange interaction that is inert in
baryons~\cite{Germani:2024miu}. Second, the scheme is economical and predictive.
Rather than pairing flavor with spin, as in the compact diquark and
Born--Oppenheimer treatments~\cite{Maiani:2023nwj}, it pairs color with flavor and
lets Fermi statistics dictate the spin content; the working assumption that
the narrow states are dominantly color $\mathbf 8 \otimes \mathbf 8$
configurations, with any $\mathbf 1\otimes \mathbf 1$ admixture
negligible~\cite{Germani:2024miu}, then leaves only a handful of exchange couplings,
all fixed from the measured spectrum. The same construction accounts at once
for the kaon-associated ($P$) and antiproton-associated ($\tilde P$) states
through the two allowed color-flavor symmetries---a unification that the
molecular and compact-diquark descriptions do not share, and which carries
over unchanged to the doubly-strange sector studied here.

There is, moreover, a specific reason to single out the doubly-strange sector
rather than treat it as a routine extrapolation. It is the one place where the
light-cloud mechanism produces a qualitatively new pattern: as the second
strange quark switches off most of the symmetric repulsion, the upper two
kaon-associated states collapse into a near-degenerate pair, a feature absent
from the well-separated non-strange and singly-strange triplets and from the
antiproton-associated series. Together with the sparse spectrum the scheme
predicts---two production-defined triplets, only $\tfrac12^-$ and $\tfrac32^-$,
with fixed internal spacings---this is precisely the structure that
distinguishes baryo-charmonium from the threshold-pinned, mixed-parity poles of
the molecular picture and from the normal hierarchy of the compact-diquark one.
Because the extension introduces no new fitted parameters, this pattern is a
parameter-free prediction of a scheme calibrated entirely on the lighter
states, so the doubly-strange sector is not merely accessible but is the
cleanest available test of the picture itself.

Although possible pentaquark configurations had been explored theoretically long before these experimental discoveries, the recent observations have renewed considerable interest in their study. As a result, pentaquark states have been investigated extensively within a variety of theoretical frameworks. Their spectroscopic properties, internal structures, interactions, and quantum-number assignments have been analyzed in detail to gain a deeper understanding of their underlying dynamics and constituent organization. Given the growing experimental capabilities and the increasing number of observed exotic hadrons, it is reasonable to anticipate the discovery of additional pentaquark states in the near future. In particular, states with quark compositions different from those identified so far are expected to emerge, providing new opportunities to probe the dynamics of multiquark systems and further test theoretical models of exotic hadron structure.

The possibility of hidden-charm pentaquarks containing multiple strange quarks has been investigated within a variety of theoretical frameworks. One of the earliest studies was performed in Ref.~\cite{Anisovich:2015zqa}, where strange and double-strange hidden-charm pentaquarks were described as diquark--diquark--antiquark systems. Within a constituent diquark model, the spin--isospin structure and mass spectra of these states were analyzed. Hidden-charm pentaquarks with triple strangeness were later studied in Ref.~\cite{Meng:2019fan}, while double-strange hidden-charm molecular pentaquarks were explored using the one-boson-exchange model in Ref.~\cite{Wang:2020bjt}. The hadro-quarkonium interpretation of double-strange pentaquarks was subsequently examined in Refs.~\cite{Ferretti:2020ewe,Ferretti:2021zis}, and doubly-strange hidden-charm molecular pentaquarks were analyzed within a meson--baryon molecular framework based on the one-boson-exchange model in Ref.~\cite{Wang:2021hql}. A unitarized coupled-channel approach constrained by the local hidden gauge formalism and heavy-quark spin symmetry predicted four double-strange hidden-charm molecular pentaquarks dominated by $\bar{D}\Omega_c^{(*)}$ and $\bar{D}_s\Xi_c^{(*)}$ components, whereas no dynamically generated resonances were found in the corresponding triple-strange sector~\cite{Roca:2024nsi}. An off-shell coupled-channel analysis predicted eight double-strangeness hidden-charm pentaquark states with quantum numbers $J^P=1/2^\pm$, $3/2^\pm$, and $5/2^-$~\cite{Clymton:2025zer}. Several of these states were found below the relevant meson--baryon thresholds and were suggested as promising candidates for future observation in the $J/\psi,\Xi$ channel. The prospects for observing hidden-charm doubly-strange pentaquarks in the decays $\Lambda_b\to J/\psi\,\Xi^-K^+$ and $\Xi_b\to J/\psi\,\Xi^-\pi^+$ were investigated within a coupled-channel unitary approach, where the predicted $P_{css}$ states emerge dynamically from vector--baryon interactions and could become visible with improved experimental mass resolution \cite{Roca:2025zyi}. The electromagnetic properties of doubly-strange hidden-charm pentaquarks have also been investigated using light-cone QCD sum rules~\cite{Ozdem:2022iqk,Ozdem:2023htj,Ozdem:2024suc} and constituent quark models~\cite{Mutuk:2024ach}.

Double strangeness constitutes a natural and largely unexplored extension of the hidden-charm pentaquark sector. Existing theoretical predictions place the masses of the doubly-strange hidden-charm pentaquarks $P_{css}$ ($c\bar{c}ssq$) in the range of approximately $4.4$--$4.8$~GeV, close to the $\Xi_c^{(\prime,*)}\bar D_s^{(*)}$ and $\bar D^{(*)}\Omega_c^{(*)}$ thresholds~\cite{Wang:2020bjt,Marse-Valera:2022khy,Roca:2024nsi,Clymton:2025zer,Marse-Valera:2024apc,Ortega:2022uyu,Wang:2025pjt}. From the experimental perspective, the search for such states has become increasingly timely. Recently, the CMS and LHCb Collaborations reported the $S=-2$ decay channels $\Lambda_b^0\to J/\psi,\Xi^-K^+$ and $\Xi_b^0\to J/\psi,\Xi^-\pi^+$~\cite{CMS:2024vnm,LHCb:2025lhk}, providing direct access to the $J/\psi,\Xi$ invariant-mass spectrum in which a doubly-strange hidden-charm pentaquark signal may appear.

We provide here the baryo-charmonium prediction for the doubly-strange sector.
The construction introduces no new fitted parameters beyond those of the
non-strange and singly-strange sectors: the spin splittings are controlled
entirely by the light cloud, so the strange--strange couplings follow from a
second application of the chromomagnetic scaling, and the overall mass scale
from an additive strange-mass increment fixed by the $P_c\!\to\!P_{cs}$ step. This is
not, however, the same as ``parameter-free.'' The splittings carry no free
parameter of their own, being fixed by the $J^{qq}$ couplings and the scaling
factor $\kappa^{qs}/\kappa^{qq}$ already determined in the lighter sectors; the
absolute scale rests in addition on the additivity ansatz, the one assumption
we cannot test directly against data. Nor does the absence of fitted parameters
imply full control of the dynamics: the scheme assumes dominance of the compact
$\mathbf 8\otimes \mathbf8$ configuration but does not fix its coupling to color-singlet
($\mathbf 1\otimes \mathbf1$) channels, an uncontrolled effect we can only estimate and carry
as a systematic. We obtain two triplets of negative-parity states,
identify a near-degenerate upper doublet in the kaon-associated series that
sets the doubly-strange sector apart from the lighter ones---within which the
level ordering formally inverts, though by a margin too small to be significant
within the model's own error budget---and check the additive increment against an
independent estimate. 

The paper is organized as follows.
Section~\ref{sec:model} reviews the model and the exchange interaction;
Sec.~\ref{sec:couplings} fixes the couplings and the strangeness scaling;
Sec.~\ref{sec:spectrum} presents the spectrum and its $J^P$ content;
Sec.~\ref{sec:check} compares with other approaches; Sec.~\ref{sec:prod}
addresses production and decay; and Sec.~\ref{sec:summary} summarizes the results of this work.

% ====================================================================
\section{The model and the exchange interaction}
\label{sec:model}
% ====================================================================
Following Ref.~\cite{Germani:2024miu}, the three light quarks form a color octet
$(qqq)_8$ and a flavor octet, bound to a color-octet $(c\bar c)_8$ pair into
an overall color singlet. Being identical fermions, the light quarks must
carry a totally antisymmetric wave function in the product of color, flavor,
spin, and orbital labels. Requiring the color and flavor parts both to
transform in the adjoint ($\mathbf 8$) representation leaves two inequivalent
three-quark tensors: one, $S$, symmetric and one, $A$, antisymmetric under a
simultaneous color-flavor exchange of a pair. Fermi statistics then fixes the
spin-orbital symmetry that must accompany each, and the two are realized
physically as the two production patterns seen in the data---states produced
in association with a kaon (class $S$, denoted $P$) and those produced in association with
an antiproton (class $A$, denoted $\tilde P$).

Within a given class the light quarks interact through the color-spin
exchange potential~\cite{Germani:2024miu}
\begin{equation}
  V \;=\; -\sum_{\rm pairs} J_{ab}\left(\tfrac12 + 2\,\bm S_a\!\cdot\!\bm S_b\right),
  \label{eq:V}
\end{equation}
where $J_{ab}$ is the exchange integral of the pair $(a,b)$ and the sum runs
over the three light-quark pairs. Its sign is set by the color state of the
pair: a color-antisymmetric ($\bar{\mathbf 3}$, attractive) pair has $J_A<0$,
a color-symmetric ($\mathbf 6$, repulsive) pair has $J_S>0$, and one-gluon
exchange relates the two by $J_S\simeq-\tfrac12 J_A$. In ordinary baryons the
three light quarks are fully color-antisymmetric, the three couplings
coincide, and the spin splitting vanishes; the pentaquark cloud instead
admits unequal $J_{ab}$ and hence a nontrivial spectrum.

Diagonalizing Eq.~\eqref{eq:V} in the space of three spin-$\tfrac12$ quarks
yields one $J=\tfrac32$ level and two $J=\tfrac12$ levels, displaced from the
spin-averaged mass by~\cite{Germani:2024miu}
\begin{align}
  \Delta E_{1/2} &= \pm\sqrt{\,J_a^2+J_b^2+J_c^2-J_aJ_b-J_bJ_c-J_cJ_a\,},
  \label{eq:dE12}\\[2pt]
  \Delta E_{3/2} &= \mp\,(J_a+J_b+J_c),
  \label{eq:dE32}
\end{align}
where $(J_a,J_b,J_c)$ denote the three pair couplings and the upper (lower)
signs hold for class $S$ ($A$), reflecting the opposite spin-orbital symmetry
imposed by Fermi statistics. A physical state then has mass $M_0+\Delta E$,
with $M_0$ the degenerate mass the triplet would carry if the exchange
interaction were switched off. The key structural point is that $M_0$ absorbs
the entire heavy-core and binding contribution, whereas
Eqs.~\eqref{eq:dE12}--\eqref{eq:dE32} depend only on the light-quark
couplings; this factorization is what allows the different strangeness sectors
to be related to one another. Throughout we take $S_{c\bar c}=0$ and an
$S$-wave core--cloud configuration, as in Ref.~\cite{Germani:2024miu}; the resulting
$J^P$ content is discussed at the end of Sec.~\ref{sec:spectrum}.

In carrying this framework into the doubly-strange sector we also inherit its
central dynamical assumption---that the physical states are dominantly the
compact color-octet $(c\bar c)_8(qqq)_8$ configuration, with the color-singlet
($\mathbf 1\otimes \mathbf1$) hadronic admixture a perturbation. This assumption is not
automatic in a new flavor sector and deserves explicit scrutiny rather than
inheritance by default. At the structural level it transfers cleanly: the color
forces that bind the cloud to the core are flavor-blind, so the color-octet
assignment of the $ssq$ cloud is identical to that of the $uud$ and $uds$
clouds, and only the chromomagnetic couplings and the baseline mass respond to
strangeness---both of which we treat explicitly in
Sec.~\ref{sec:couplings}. At the dynamical level its validity is
genuinely sector-specific: it is favored by the increased compactness of the
heavier cloud but challenged by the greater proximity of the doubly-strange
states to the open-charm thresholds. We defer the quantitative version of this
discussion, and the size of the resulting systematic, to Sec.~\ref{sec:check},
flagging here only that the $\mathbf 8\otimes \mathbf8$ dominance---and not the kinematics or
the spin algebra---is the assumption that most directly distinguishes the
present scheme from the molecular interpretations.

% ====================================================================
\section{Light-quark couplings and strangeness scaling}
\label{sec:couplings}
% ====================================================================
The non-strange couplings are fixed by the three $P_c$ masses. Solving
Eqs.~\eqref{eq:dE12}--\eqref{eq:dE32} for the $uud$ cloud, in which the
identical $u$ quarks sit in the symmetric pairing and the $ud$ pairs in the
antisymmetric one, gives~\cite{Germani:2024miu}
\begin{align}
  J_S^{qq}&=29.9^{+2.5}_{-2.8}\ \text{MeV},\notag\\
  J_A^{qq}&=-42.8^{+2.4}_{-1.6}\ \text{MeV}.
  \label{eq:Jqq}
\end{align}
Strangeness enters through the chromomagnetic hierarchy
$\kappa_{ij}\propto 1/(m_i m_j)$. With $\kappa^{qs}/\kappa^{qq}\simeq0.6$
known from the baryon spectrum~\cite{Ali:2019roi}, one obtains
$J^{qs}=0.6\,J^{qq}$, reproducing the singly-strange sector. Applying the
same scaling a second time, $\kappa^{ss}/\kappa^{qs}\simeq\kappa^{qs}/\kappa^{qq}
\simeq0.6$, fixes the strange--strange couplings needed here,
\begin{align}
  J_S^{ss}&=0.36\,J_S^{qq}=10.8^{+0.9}_{-1.0}\ \text{MeV},\notag\\
  J_A^{ss}&=0.36\,J_A^{qq}=-15.4^{+0.9}_{-0.6}\ \text{MeV}.
  \label{eq:Jss}
\end{align}
Here, no additional fit is involved: $J^{ss}$ follows from
Eq.~\eqref{eq:Jqq} and the single scaling factor. The second application is
not an independent assumption but the exact consequence of the
$\kappa_{ij}\propto1/(m_im_j)$ form: with multiplicative constituent masses
$\kappa^{qs}/\kappa^{qq}=\kappa^{ss}/\kappa^{qs}=m_q/m_s$, so the same ratio
governs both steps. The genuine approximation lies elsewhere, and it is the
import of the ratio itself. The chromomagnetic coupling is not purely kinematic
but carries a short-distance overlap factor---the contact probability
$\propto|\psi_{ij}(0)|^2$---and $\kappa^{qs}/\kappa^{qq}\simeq0.6$ is extracted
in the color-singlet baryon environment, whereas the cloud here is a color
octet whose spatial wave function need not coincide. Three points bound the
resulting uncertainty. First, the baryon-extracted ratio agrees with the
constituent-mass ratio $m_q/m_s\simeq0.6$, indicating that the flavor
dependence of the overlap is mild even in baryons, the kinematic factor
dominating. Second, what enters the splittings is the ratio of
couplings, in which a flavor-blind change of the overlap between the singlet
and octet environments---a common rescaling of $|\psi(0)|^2$ for all
pairs---cancels exactly; only a flavor-dependent difference in the octet
overlap survives, and that is a second-order effect. Third, and most directly,
the same ratio applied within this octet-cloud framework already reproduces the
observed singly-strange $P_{cs}$ states~\cite{Germani:2024miu}, so its validity
in the exotic environment is corroborated by pentaquark data for the first
strange substitution; the doubly-strange case is the extrapolation of a rule
already tested once in the relevant setting. We do not derive the octet overlap
from first principles, and any residual departure---from whatever source---is
subsumed in the conservative $0.5$--$0.7$ scan of $\kappa^{qs}/\kappa^{qq}$
quantified in Sec.~\ref{sec:check}, under which the splittings move by under
$\sim10$~MeV and the qualitative pattern is unchanged. The intermediate
singly-strange couplings, which enter the $A$-class construction below, are
$J_S^{qs}=17.9^{+1.5}_{-1.7}$ and $J_A^{qs}=-25.7^{+1.4}_{-1.0}$~MeV; the
three-flavor $uds$ cloud additionally requires the mixed value
$J^{ds}=\tfrac12(1+k)\,J_S^{qs}\simeq-3.9$~MeV, with $k=J_A/J_S\simeq-1.43$,
but the doubly-strange cloud does not, as explained in
Sec.~\ref{sec:spectrum}. The complete set of exchange couplings is collected
in Table~\ref{tab:couplings}.

\begin{table}[h]
\caption{Light-quark exchange couplings (MeV). $J_S$ ($J_A$) is the coupling
of a color-symmetric (antisymmetric) pair. The $qq$ values are fitted to the
$P_c$ masses~\cite{Germani:2024miu}; the $qs$ and $ss$ values follow from the
chromomagnetic scaling $\kappa_{ij}\propto 1/(m_im_j)$ with
$\kappa^{qs}/\kappa^{qq}\simeq0.6$. The doubly-strange spectrum uses the $qs$
and $ss$ entries.}
\label{tab:couplings}
\begin{ruledtabular}
\begin{tabular}{lcc}
Pair & $J_S$ (MeV) & $J_A$ (MeV) \\
\hline
$qq$ & $29.9^{+2.5}_{-2.8}$ & $-42.8^{+2.4}_{-1.6}$ \\
$qs$ & $17.9^{+1.5}_{-1.7}$ & $-25.7^{+1.4}_{-1.0}$ \\
$ss$ & $10.8^{+0.9}_{-1.0}$ & $-15.4^{+0.9}_{-0.6}$ \\
\end{tabular}
\end{ruledtabular}
\end{table}

The baselines are likewise extrapolated. The non-strange and
singly-strange degenerate masses differ by a strange-mass increment
\begin{equation}
  \Delta_s \equiv M_0^{(s)}-M_0 = 127^{+6}_{-5}\ \text{MeV},
\end{equation}
extracted directly from the observed states. Assuming additivity, the
doubly-strange baseline is
\begin{equation}
  M_0^{(ss)} = M_0 + 2\,\Delta_s ,
  \label{eq:M0ss}
\end{equation}
for each production class. We test this additivity a posteriori in
Sec.~\ref{sec:check}. A single, class-independent $\Delta_s$ is applied to
both series, which differ only through their non-strange baselines $M_0^{S}$
and $M_0^{A}$ inherited from Ref.~\cite{Germani:2024miu}; Eq.~\eqref{eq:M0ss}
then fixes the doubly-strange baselines quoted in Sec.~\ref{sec:spectrum}
($M_0^{(ss,S)}=4639$ and $M_0^{(ss,A)}=4514$~MeV), corresponding to non-strange
values $M_0^{S}\simeq4385$ and $M_0^{A}\simeq4260$~MeV. The two baselines are
not on an equal empirical footing: $M_0^{S}$ is anchored directly to the three
observed $P_c$ masses, whereas no non-strange antiproton-associated
($\tilde P$) state has been observed, so $M_0^{A}$ is inferred indirectly
within Ref.~\cite{Germani:2024miu} (through the strange $A$-class
$P_{\psi s}^{\Lambda}(4338)$ with the strange-mass increment undone). The $A$-class
doubly-strange predictions therefore rest on a longer extrapolation chain than
the $S$-class ones and should be read with correspondingly greater caution.

% ====================================================================
\section{The doubly-strange spectrum}
\label{sec:spectrum}
\begin{figure}[h!]
\centering
\resizebox{\columnwidth}{!}{%
\begin{tikzpicture}[x=1cm,y=1cm]
  \draw[->] (0,0) -- (0,13.0);
  \foreach \M/\Y in {4450/2.0,4500/4.0,4550/6.0,4600/8.0,4650/10.0,4700/12.0}{
     \draw (-0.1,\Y)--(0,\Y);
     \node[left,scale=0.7] at (-0.16,\Y){\M};}
  \node[rotate=90,scale=0.8] at (-1.0,6.4) {Mass (MeV)};
  \foreach \Y/\lab in {0.60/{$J/\psi\,\Xi$},1.48/{$\Xi_c\bar D_s$},
                       5.88/{$\Xi_c'\bar D_s$},7.24/{$\Xi_c\bar D_s^*$},
                       8.56/{$\Xi_c^*\bar D_s$}}{
     \draw[black!70,dashed] (0.3,\Y)--(6.6,\Y) node[right,black!70,scale=0.6]{\lab};}
  \foreach \Y/\j in {3.24/{$\tfrac12^-$},5.36/{$\tfrac32^-$},5.88/{$\tfrac12^-$}}{
     \draw[very thick,blue] (1.1,\Y)--(2.3,\Y);
     \node[blue,scale=0.6,right] at (2.32,\Y){\j};}
  \node[blue,scale=0.75] at (1.7,2.7) {$\tilde P_{css}\,(A)$};
  \draw[very thick,red] (4.5,8.12)--(5.7,8.12);
  \node[red,scale=0.6,right] at (5.72,8.12){$\tfrac12^-$};
  \draw[very thick,red] (4.5,11.0)--(5.7,11.0);
  \draw[very thick,red] (4.5,11.16)--(5.7,11.16);
  \draw[red,thin] (5.7,11.0)--(6.05,10.7);
  \node[red,scale=0.6,right] at (6.08,10.7){$\tfrac12^-$};
  \draw[red,thin] (5.7,11.16)--(6.05,11.46);
  \node[red,scale=0.6,right] at (6.08,11.46){$\tfrac32^-$};
  \node[red,scale=0.75] at (5.1,7.6) {$P_{css}\,(S)$};
\end{tikzpicture}}
\caption{Predicted negative-parity doubly-strange pentaquark levels:
$A$ class (blue, left) and $S$ class (red, right). Dashed lines mark
the relevant two-hadron thresholds.}
\label{fig:levels}
\end{figure}
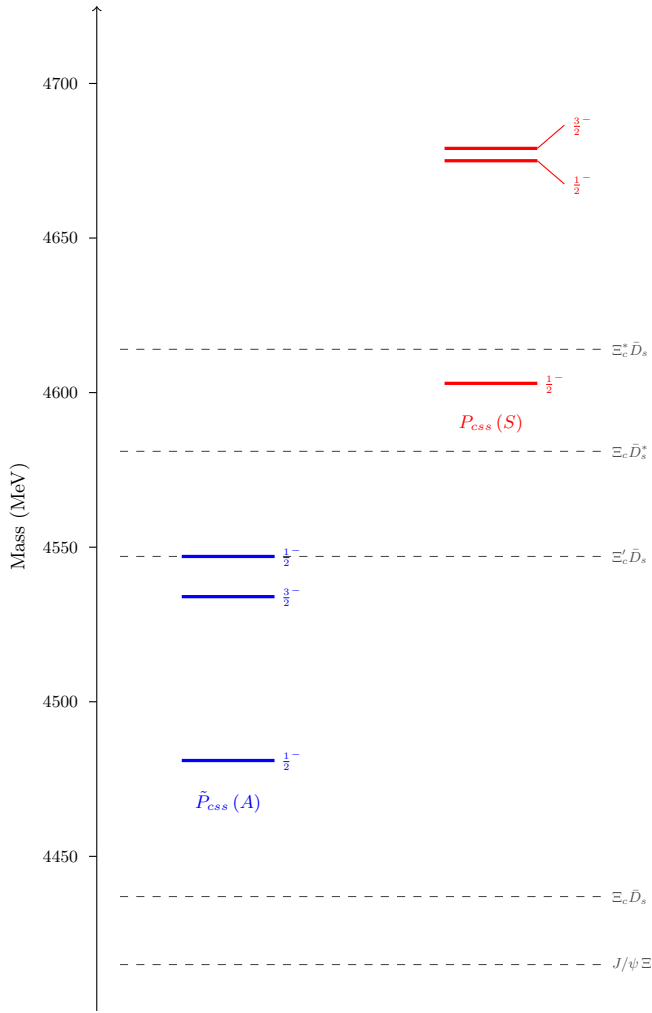

% ====================================================================
The light content is $ssq$ (the $uss$ and $dss$ states form an isospin
doublet, degenerate in this approximation); the neutral $P_{css}^0$
($c\bar c ssu$) and the charged $P_{css}^-$ ($c\bar c ssd$) decay to
$J/\psi\,\Xi^0$ and $J/\psi\,\Xi^-$, respectively. The electromagnetic and
isospin-breaking splitting between the two charge states is expected at the
few-MeV level---itself below the $\sim4$~MeV gap of the upper $S$-class pair
discussed below---and is neglected throughout. Because two quarks are
identical, $ssq$ maps directly onto the $uud$ template, with no mixed-pair
averaging of the kind required for the three-flavor $uds$ cloud. In the
$S$ class the $ss$ pair occupies the symmetric repulsive slot, $J_S^{ss}$,
and the two $qs$ pairs are antisymmetric, $J_A^{qs}$; in the $A$ class the
roles reverse ($qs$ symmetric $J_S^{qs}$, $ss$ antisymmetric $J_A^{ss}$).
Explicitly, the color-flavor tensors of the $ssq$ cloud, built in analogy with
the $uud$ and $\Lambda$-type constructions of Ref.~\cite{Germani:2024miu}, are
\begin{align}
  S^{ssq}_{\alpha\beta\gamma}&=s_{[\alpha}q_{\beta]}s_{\gamma}
        +s_{[\gamma}q_{\beta]}s_{\alpha},\label{eq:Sssq}\\
  A^{ssq}_{\alpha\beta\gamma}&=s_{(\alpha}q_{\beta)}s_{\gamma}
        -s_{(\gamma}q_{\beta)}s_{\alpha},\label{eq:Assq}
\end{align}
where square brackets and parentheses denote anti-symmetrization and
symmetrization of the enclosed color indices. Both tensors carry the
mixed-symmetry flavor structure of the octet ($\Xi$-type) $ssq$ baryon; the
fully symmetric decuplet ($\Xi^*$-type) $ssq$ combination is excluded at the
outset by the requirement of Sec.~\ref{sec:model} that the flavor part
transform in the adjoint $\mathbf 8$ and not the $\mathbf{10}$, so no
flavor-decuplet admixture enters the construction. In $S^{ssq}$ the two $s$ quarks
($\alpha,\gamma$) form the symmetric, repulsive pairing $J_S^{ss}$ and each
$sq$ pair the antisymmetric, attractive one $J_A^{qs}$; in $A^{ssq}$ the
assignment reverses. Because two quarks are identical, the doubly-strange
cloud maps onto the $uud$ template under $u\to s$, $d\to q$ with a unique pair
assignment, and the mixed-pair averaging required for the three-flavor $uds$
cloud does not arise. With Eqs.~\eqref{eq:dE12}--\eqref{eq:dE32} the
splittings are
\begin{align}
  S:&\quad \Delta E_{1/2}=36.3^{+1.4}_{-1.6},\quad
        \Delta E_{3/2}=40.5^{+2.3}_{-2.9}\ \text{MeV},\notag\\
  A:&\quad \Delta E_{1/2}=33.2^{+1.7}_{-1.8},\quad
        \Delta E_{3/2}=20.6^{+3.1}_{-3.4}\ \text{MeV}.
\end{align}
These are evaluated from the unrounded couplings $J^{qs}=0.6\,J^{qq}$ and
$J^{ss}=0.36\,J^{qq}$; reinserting the rounded entries of
Table~\ref{tab:couplings} reproduces them only to $\approx0.1$--$0.2$~MeV, well
inside the quoted uncertainties.
Combined with the baselines $M_0^{(ss,S)}=4639^{+12}_{-8}$ and
$M_0^{(ss,A)}=4514^{+6}_{-6}$~MeV, this gives the spectrum in
Table~\ref{tab:spectrum} and Fig.~\ref{fig:levels}. We stress that these
absolute masses are conditional on the additivity ansatz of
Eq.~\eqref{eq:M0ss}: the baselines $M_0^{(ss)}=M_0+2\Delta_s$ carry the entire
scale, so a moderate ($10$--$20\%$) departure from additivity---which the
consistency check of Sec.~\ref{sec:check} does not exclude---would shift all
six masses rigidly by up to several tens of MeV. The uncertainties quoted in
Table~\ref{tab:spectrum} are the coupling-fit and $\Delta_s$-extraction errors
only; the realistic absolute uncertainty, including the additivity and
color-singlet-mixing systematics, is the larger $\sim\pm55$~MeV of
Table~\ref{tab:budget}. What is not conditional on additivity is the
internal structure---the fixed internal spacings and the near-degenerate
kaon-associated doublet---which a rigid common shift of the scale leaves
intact.

\begin{table}[!h]
\caption{Predicted doubly-strange hidden-charm pentaquark masses (MeV).
$S$ denotes the kaon-associated class, $A$ the antiproton-associated
($\tilde P$) class. States are listed in order of increasing mass. The two
$J=\tfrac12$ members of each triplet lie symmetrically at
$M_0\pm\Delta E_{1/2}$ and hence carry identical uncertainties; the
$J=\tfrac32$ member, at $M_0+\Delta E_{3/2}$, differs slightly. The quoted
uncertainties are the coupling-fit and $\Delta_s$-extraction errors only; the
absolute scale carries an additional $\sim\pm55$~MeV systematic
(Table~\ref{tab:budget}), and all values are conditional on the additivity
ansatz of Eq.~\eqref{eq:M0ss}.}
\label{tab:spectrum}
\begin{ruledtabular}
\begin{tabular}{lcc}
State & $J^P$ & Mass \\
\hline
$P_{css}$        & $\tfrac12^-$ & $4603^{+12}_{-8}$ \\
$P_{css}'$       & $\tfrac12^-$ & $4675^{+12}_{-8}$ \\
$P_{css}''$      & $\tfrac32^-$ & $4679^{+12}_{-9}$ \\[3pt]
$\tilde P_{css}$  & $\tfrac12^-$ & $4481^{+6}_{-6}$ \\
$\tilde P_{css}'$ & $\tfrac32^-$ & $4534^{+7}_{-7}$ \\
$\tilde P_{css}''$& $\tfrac12^-$ & $4547^{+6}_{-6}$ \\
\end{tabular}
\end{ruledtabular}
\end{table}

The salient qualitative feature is the collapse of the upper two
$S$-class states into a near-degenerate pair. For the non-strange and
singly-strange triplets one has
$\Delta E_{3/2}<\Delta E_{1/2}$, so the $\tfrac32^-$ state lies between the
two $\tfrac12^-$ states and the pattern reads $\tfrac12,\tfrac32,\tfrac12$
with increasing mass, the upper $\tfrac12^-$ well separated from the $\tfrac32^-$.
In the doubly-strange $S$ class the small positive
$J_S^{ss}$ no longer dominates the sizable negative $J_A^{qs}$, and
$\Delta E_{3/2}$ rises to within a few MeV of $\Delta E_{1/2}$, so the upper
$\tfrac12^-$ and the $\tfrac32^-$ become nearly degenerate; formally
$\Delta E_{3/2}>\Delta E_{1/2}$, placing the $\tfrac32^-$ marginally on top and
giving the ordering $\tfrac12,\tfrac12,\tfrac32$. From
Eqs.~\eqref{eq:dE12}--\eqref{eq:dE32} the crossover is governed by a compact
condition,
\begin{equation}
  \Delta E_{3/2}>\Delta E_{1/2}\ \Longleftrightarrow\
  \left|\frac{J_A^{qq}}{J_S^{qq}}\right| > 2\,\frac{\kappa^{qs}}{\kappa^{qq}},
  \label{eq:crossover}
\end{equation}
i.e.\ $|J_A^{qq}/J_S^{qq}|>1.2$ for the adopted scaling. The fitted ratio
is $1.43^{+0.10}_{-0.10}$, so the inversion is realized; sampling the fitted
couplings $J_{S,A}^{qq}$ from (asymmetric) Gaussians with the errors of
Table~\ref{tab:couplings} and the ratio $\kappa^{qs}/\kappa^{qq}$ uniformly
over the conservative band $[0.5,0.7]$, the inversion condition
$|J_A^{qq}/J_S^{qq}|>2\,\kappa^{qs}/\kappa^{qq}$ is satisfied in $96\%$ of
samples, the failures concentrated near the upper edge $\kappa^{qs}/\kappa^{qq}\!\to\!0.7$
where the threshold $2\kappa$ approaches the central ratio $1.43$.
The crossover into near-degeneracy is therefore a robust feature of the
scheme: should $\kappa^{qs}/\kappa^{qq}$ be appreciably larger than $0.72$, the
upper pair would instead separate in the normal order. The $A$ class retains
the standard well-separated $\tfrac12,\tfrac32,\tfrac12$ ordering. We stress,
however, that two claims of very different robustness are bundled here. The
robust, resolution-independent prediction is that the two upper $S$-class
states form a near-degenerate doublet---the $\tfrac12^-$--$\tfrac32^-$
spacing collapsing from the $\sim40$~MeV of the lighter sectors to only
$\sim4$~MeV. The strict $\tfrac32^-$-on-top ordering within that doublet
is a finer matter, and assessing it requires the appropriate error bar. The
$\sim13$~MeV figure of Table~\ref{tab:budget} is the systematic on an
individual splitting, whereas the inversion is the sign of the
difference $D\equiv\Delta E_{3/2}-\Delta E_{1/2}$, and the two systematics that
dominate that $\sim13$~MeV are largely common-mode in $D$. The additive-scale
uncertainty cancels in any splitting difference identically; and the
chromomagnetic scan---the largest splitting systematic---shifts $\Delta E_{3/2}$
and $\Delta E_{1/2}$ coherently, so that
$D=(\kappa^{qs}/\kappa^{qq})\,|J_A^{qq}|-2(\kappa^{qs}/\kappa^{qq})^2 J_S^{qq}$
remains positive across the entire conservative band, falling from
$6.5$~MeV at $\kappa^{qs}/\kappa^{qq}=0.5$ to $0.7$~MeV at $0.7$ and crossing
zero only at $\kappa^\star=0.716$ [Eq.~\eqref{eq:crossover}]: the inversion
direction is stable over the scan even as each level moves by $\pm8$~MeV.
Comparing the $\sim4$~MeV margin to the $\pm13$~MeV per-level error therefore
overstates the uncertainty. The relevant quantity is the error on $D$, into
which only the $J^P$-differential systematics propagate---chiefly the
channel-dependent $\mathbf 8\otimes \mathbf 8$--$\mathbf 1\otimes \mathbf1$ mixing, itself of order a few MeV.
Propagating the coupling and $\kappa$ uncertainties into $D$ directly, the
Monte Carlo above returns $D>0$ in $96\%$ of samples, an internal significance
of $\approx1.8\sigma$ for the ordering, with the failures confined to the
$\kappa\!\to\!0.72$ edge. This is the correct measure, and it places the
inversion as a $\sim2\sigma$ preference---firmer than the naive ratio of the
$\sim4$~MeV margin to the $\sim13$~MeV per-level systematic ($\approx0.3$) would
suggest, but short of a sharp prediction, and in any case below present
$J/\psi\,\Xi$ resolution. The near-degeneracy is untouched by all of this: it is
the $\sim4$~MeV magnitude of $D$, robust across the scan. The doublet is
thus the firm prediction, and its internal ordering a $\sim2\sigma$ tendency for
a future high-resolution spin analysis to settle.

This inequality has a transparent physical reading. The $\tfrac32^-$ level is
held down by the repulsive, symmetric pair, which enters
$\Delta E_{3/2}=-(J_a+J_b+J_c)$ [Eq.~\eqref{eq:dE32}] with the sign opposite to
the two attractive, antisymmetric pairs; the ordering stays normal only while
that repulsion is strong enough, $|J_{\rm antisym}|<2\,J_{\rm sym}$. In the $S$
class the symmetric pair is the $ss$ pair and the antisymmetric ones are $qs$
[Eqs.~\eqref{eq:Sssq}--\eqref{eq:Assq}]. Because the chromomagnetic coupling
scales as $1/(m_im_j)$, placing both quarks of the symmetric pair in the
strange sector suppresses it by $\kappa^2\simeq0.36$, while each attractive
$qs$ pair is weakened only by $\kappa\simeq0.6$. The second strange quark thus
acts selectively on the one coupling that enforces the normal ordering: it
switches off most of the symmetric repulsion, lowering the threshold ratio from
$|J_A^{qq}/J_S^{qq}|>2$ in the non-strange triplet to $>2\kappa\simeq1.2$ here.
The light-cloud ratio $1.43$ lies between these two values, so the level scheme
that is normal for $P_c$ and $P_{cs}$ inverts for $P_{css}$. The effect is
specific to the second strange quark and to the $S$ class: with fewer strange
quarks the symmetric pair is never a pure $ss$ pair, and in the $A$ class the
assignment reverses---there the $ss$ pair is the antisymmetric, attractive one
($\propto\kappa^2$) while the repulsive pairs are $qs$ ($\propto\kappa$), so the
attraction is suppressed more than the repulsion and the standard ordering
survives.

The $J^P$ content of these states follows from the $S$-wave, $S_{c\bar c}=0$
assumption. The pentaquark parity is $P=(-1)(+1)(-1)^{L}=(-1)^{L+1}$, where
$-1$ and $+1$ are the intrinsic parities of the $c\bar c$ pair and the
three-quark cloud and $L$ is their relative orbital angular momentum; with
$L=0$ one has $P=-1$, and because the exchange interaction acts only on the
light spins the accessible quantum numbers are exactly $\tfrac12^-$ and
$\tfrac32^-$. This is the same set realized by the observed $P_c$ and $P_{cs}$
states---the $P_{\psi s}^{\Lambda}(4338)$ has measured $J^P=\tfrac12^-$---and
the relevant one for $S$-wave hadronic thresholds. Other quantum numbers
require structure beyond the minimal scheme: positive parity arises from an
$L=1$ core--cloud excitation, which would introduce both an orbital energy gap
and the spin-orbit term consistently dropped at $L=0$, while $J=\tfrac52$
needs $S_{c\bar c}=1$ together with the associated small, currently
unconstrained hyperfine couplings. We accordingly present the negative-parity
$L=0$ multiplets as the robust prediction and leave their positive-parity and
$\tfrac52$ partners to future work.

% ====================================================================
\section{Comparison and discussion}
\label{sec:check}
% ====================================================================
The additivity of the strange-mass increment, Eq.~\eqref{eq:M0ss}, can be tested by
inverting the construction. Because $\Delta E_{1/2}$ is baseline-free, an
independently predicted $S$-class ground state $M_\star$ fixes
$M_0^{(ss)}=M_\star+\Delta E_{1/2}$, and hence an effective per-quark
increment $\Delta_s^{\rm eff}=(M_0^{(ss)}-M_0)/2$. The diquark QCD sum-rule
analysis of Ref.~\cite{Wang:2025pjt} places a $\tfrac12^-$ doubly-strange
level at $M_\star=4.61\pm0.11$~GeV, numerically close to our $S$-class
ground state; inverting gives
\begin{equation}
  \Delta_s^{\rm eff}=131\pm55\ \text{MeV},
\end{equation}
consistent with the singly-strange $\Delta_s=127^{+6}_{-5}$~MeV. The central
values coincide to within a few MeV, although the $\pm55$~MeV spread makes this
a weak constraint that does not exclude moderate departures from additivity,
and the comparison is against another theoretical estimate rather than data.
This is, however, not the only handle on additivity, and the sturdier one is
data-driven. The physical content of the ansatz---that each strange
substitution shifts the baseline by a nearly constant amount---is an
experimentally established regularity of the very light-baryon spectrum from
which the couplings are drawn: the decuplet $\Delta(1232)$, $\Sigma^*(1385)$,
$\Xi^*(1530)$, $\Omega(1672)$ is equally spaced to within $\sim10$~MeV,
 a measured fact rather than a model output. What is checked only against
theory is therefore the narrower statement that this empirical additivity
transfers with the same coefficient to the doubly-strange pentaquark
baseline---not additivity itself, which is anchored in data. A direct test in
the pentaquark sector is moreover within reach: because the splittings are
baseline-free, a single measured doubly-strange mass fixes $M_0^{(ss)}$, and
with it any departure from additivity, outright---turning the ansatz into a
measured quantity. The $S=-2$ $J/\psi\,\Xi$ analyses now underway
(Sec.~\ref{sec:prod}) are exactly what would supply it.
It should be stressed that this anchor cuts only one way: Ref.~\cite{Wang:2025pjt}
constrains our absolute scale through its ground-state mass, but it
returns the normal spin hierarchy and so contradicts our
headline inversion. The coincidence of one mass therefore validates the scale,
not the splitting structure, and the two references we lean on for the scale
and for the ordering test are not mutually consistent on the ordering itself.
With that caveat, the corresponding anchored
partners, $\tfrac32^-$ at $4687$ and $\tfrac12^-$ at $4683$~MeV, lie within
$\sim10$~MeV of the {\it ab initio} entries in Table~\ref{tab:spectrum}. The
same analysis places its lowest $\tfrac12^-$ level at $4.48$~GeV, matching our
$A$-class ground state ($4481$~MeV).

\begin{table*}[!h]
\caption{Doubly-strange hidden-charm pentaquark masses (MeV) in the present
scheme compared with molecular coupled-channel and QCD sum-rule results.
Parenthesized spin lists denote states degenerate in the corresponding
approach.}
\label{tab:compare}
\begin{ruledtabular}
\begin{tabular}{lll}
Approach & Method & States ($J^P$:\,mass) \\
\hline
This work ($S$ class) & baryo-charmonium & $\tfrac12^-$:4603, $\tfrac12^-$:4675, $\tfrac32^-$:4679 \\
This work ($A$ class) & baryo-charmonium & $\tfrac12^-$:4481, $\tfrac32^-$:4534, $\tfrac12^-$:4547 \\
Ref.~\cite{Roca:2024nsi} & molecular (unitarized) & $\tfrac12^-$:4535, $\tfrac32^-$:4602, $(\tfrac12^-,\tfrac32^-)$:4675, $(\tfrac12^-,\tfrac32^-,\tfrac52^-)$:4743 \\
Ref.~\cite{Clymton:2025zer} & molecular (off-shell coupled channel) & $\tfrac12^-$:4437,\,4504,\,4704;\ $\tfrac32^-$:4541;\ $\tfrac52^-$:4757 \\
Ref.~\cite{Marse-Valera:2022khy} & molecular (unitarized) & $\tfrac12^-$:4493;\ $(\tfrac12^-,\tfrac32^-)$:4633 \\
Ref.~\cite{Wang:2025pjt} & QCD sum rules (diquark) & $\tfrac12^-$:4480, $\tfrac32^-$:4510, $\tfrac52^-$:4540 (lowest; up to 4710) \\
\end{tabular}
\end{ruledtabular}
\end{table*}

\begin{figure*}[!t]
\centering
\resizebox{\textwidth}{!}{%
\begin{tikzpicture}[x=1cm,y=0.018cm]
  \foreach \Y/\lab in {15/{$J/\psi\,\Xi$},37/{$\Xi_c\bar D_s$},147/{$\Xi_c'\bar D_s$},181/{$\Xi_c\bar D_s^*$},214/{$\Xi_c^*\bar D_s$}}{
     \draw[black!70,dashed] (1.2,\Y)--(16.6,\Y) node[right,black!70,scale=0.6]{\lab};}
 
  \draw[->] (1.2,-25)--(1.2,425);
  \foreach \Y/\M in {0/4400,100/4500,200/4600,300/4700,400/4800}{
     \draw (1.05,\Y)--(1.2,\Y);
     \node[left,scale=0.7] at (1.0,\Y){\M};}
  \node[rotate=90,scale=0.8] at (0.4,200){Mass (MeV)};
 
  \node[align=center,scale=0.62] at (2.2,420){This work\\($S$)};
  \node[align=center,scale=0.62] at (4.9,420){This work\\($A$)};
  \node[align=center,scale=0.62] at (7.6,418){Ref.~\cite{Roca:2024nsi}};
  \node[align=center,scale=0.62] at (10.3,418){Ref.~\cite{Clymton:2025zer}};
  \node[align=center,scale=0.62] at (13.0,418){Ref.~\cite{Marse-Valera:2022khy}};
  \node[align=center,scale=0.62] at (15.7,418){Ref.~\cite{Wang:2025pjt}};

  \draw[very thick,red] (1.5,203)--(2.9,203);
  \draw[very thick,red] (1.5,275)--(2.9,275);
  \draw[very thick,red] (1.5,279)--(2.9,279);

  \draw[very thick,blue] (4.2,81)--(5.6,81);
  \draw[very thick,blue] (4.2,134)--(5.6,134);
  \draw[very thick,blue] (4.2,147)--(5.6,147);
 
  \foreach \Y in {135,202,275,343}{\draw[very thick,green!55!black] (6.9,\Y)--(8.3,\Y);}
 
  \foreach \Y in {37,104,141,304,357}{\draw[very thick,orange!85!black] (9.6,\Y)--(11.0,\Y);}
 
  \foreach \Y in {93,233}{\draw[very thick,violet] (12.3,\Y)--(13.7,\Y);}
  
  \foreach \Y in {80,90,110,140,160,200,210,250,260,290,300,310}{\draw[very thick,teal] (15.0,\Y)--(16.4,\Y);}
\end{tikzpicture}}
\caption{Comparison of doubly-strange hidden-charm pentaquark mass spectra from different theoretical approaches. Dashed horizontal lines indicate the relevant two-hadron thresholds.}
\label{fig:compare}
\end{figure*}
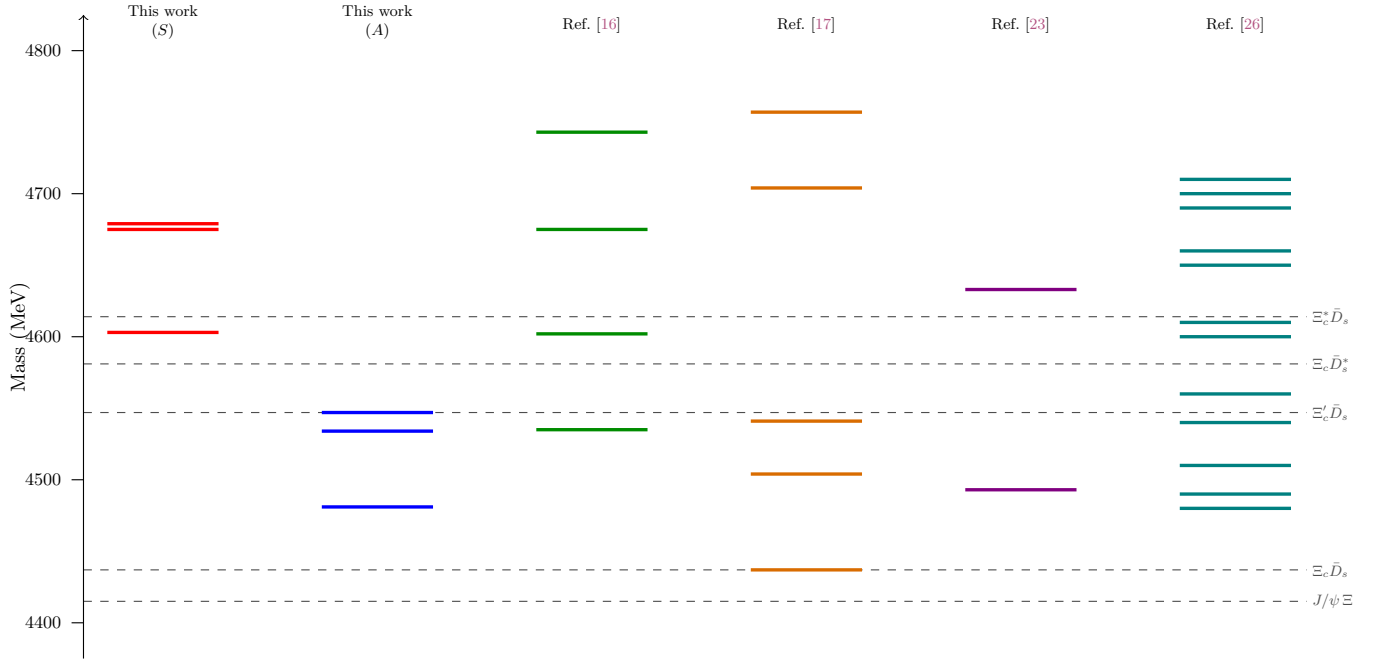

Table~\ref{tab:compare} sets our predictions beside the principal alternative
calculations of the doubly-strange sector. References~\cite{Roca:2024nsi,Marse-Valera:2022khy,Marse-Valera:2024apc}
unitarize a coupled-channel meson--baryon interaction built from vector-meson
exchange and generate the states dynamically as poles; Ref.~\cite{Clymton:2025zer} solves
an off-shell coupled-channel Bethe--Salpeter equation with a
heavy-quark-symmetric Lagrangian and, in addition, finds positive-parity
$P$-wave partners; and Ref.~\cite{Wang:2025pjt} evaluates diquark--diquark--antiquark
interpolating currents in QCD sum rules, obtaining a full
$\tfrac12^-,\tfrac32^-,\tfrac52^-$ spectrum. Despite the disparate dynamics, the predicted
mass windows largely coincide, as Fig.~\ref{fig:compare} shows. Our $A$-class triplet
($4481,4534,4547$~MeV) tracks the lowest molecular cluster: the lowest
$\tfrac12^-$ falls in the $\bar D_s\Xi_c$--$\bar D_s\Xi_c'$ region populated by
Refs.~\cite{Roca:2024nsi,Clymton:2025zer,Marse-Valera:2022khy} ($4437$--$4535$~MeV), and our $\tfrac32^-$ at
$4534$~MeV sits within a few MeV of the $4541$~MeV $\tfrac32^-$ pole of
Ref.~\cite{Clymton:2025zer}. Our $S$-class triplet ($4603$--$4679$~MeV) overlaps the upper
cluster, its lowest member coinciding with the $4602$~MeV $\tfrac32^-$
pole of Ref.~\cite{Roca:2024nsi} and the $\tfrac12^-$ sum-rule level at
$4.61$~GeV of Ref.~\cite{Wang:2025pjt},
while lying above the $J/\psi\,\Xi$ ($\sim4415$~MeV) and $\Xi_c\bar D_s$
($\sim4437$~MeV) thresholds so that both decay channels are open.

Two systematic differences are expected and informative. The molecular
analyses generate a state at each two-hadron threshold, including $\tfrac52^-$
members built on $\tfrac32^+$ baryons and, in Ref.~\cite{Clymton:2025zer}, positive-parity
$P$-wave states, and the diquark sum-rule spectrum of Ref.~\cite{Wang:2025pjt}
likewise spans $\tfrac12^-,\tfrac32^-,\tfrac52^-$; our minimal scheme
($S_{c\bar c}=0$, $S$-wave core--cloud) yields only the negative-parity
$\tfrac12^-$ and $\tfrac32^-$ multiplets. The
members absent from our spectrum are precisely those requiring the additional
structure---$S_{c\bar c}=1$ or $L=1$---discussed in Sec.~\ref{sec:spectrum}.
Moreover, the molecular masses carry a regularization-cutoff dependence (a
$\sim50$--$80$~MeV spread between $\Lambda=600$ and $800$~MeV in
Ref.~\cite{Roca:2024nsi}), whereas our values follow from data-fixed couplings and the
additive increment with no free regulator.

Beyond these multiplicities it is instructive to compare the predicted
orderings, since the inversion is offered as a discriminator. In the
molecular analyses the ordering is set by where the thresholds lie rather than
by an internal spin mechanism: a $\tfrac32^-$ pole sits above a $\tfrac12^-$ one
when it is built on a heavier hadron pair---a $\tfrac32^+$ charmed baryon or a
vector $\bar D_s^*$. References~\cite{Roca:2024nsi,Clymton:2025zer} accordingly
do place $\tfrac32^-$ states above their lowest $\tfrac12^-$, but as
threshold-pinned poles spread over $\gtrsim100$~MeV, not as a fixed-spacing
triplet with a near-degenerate $\tfrac12^-$--$\tfrac32^-$ pair at its top. The
compact diquark sum rules of Ref.~\cite{Wang:2025pjt}, by contrast, return a
normal spin hierarchy---$\tfrac12^-$ below $\tfrac32^-$ below
$\tfrac52^-$ with increasing mass---and so do not reproduce the inversion at
all. The pattern is therefore not generic: it is absent from the diquark
spectrum and appears in the molecular one only as a reflection of the threshold
ordering. What singles out the present prediction is consequently not the
inversion in isolation---a $\tfrac32^-$ lying above a $\tfrac12^-$ can arise for
unrelated reasons---but the inversion together with the rest of the
pattern: exactly two $\tfrac12^-$ and one $\tfrac32^-$ per production triplet,
fixed internal spacings independent of any threshold, a near-degenerate upper
pair, and no $\tfrac52^-$ or positive-parity partners. It is this combination,
not the ordering alone, that distinguishes the scheme from its competitors.

Two inputs control the results and warrant comment. The chromomagnetic ratio
$\kappa^{qs}/\kappa^{qq}\simeq0.6$ is taken from the light-baryon spectrum;
varying it over a conservative $0.5$--$0.7$ range moves the splittings by under $\sim\!10$~MeV (Table~\ref{tab:kappa})
and, as Eq.~\eqref{eq:crossover} makes explicit, leaves the
$S$-class inversion intact as long as the ratio stays below $0.72$. The
additive strange-mass increment of Eq.~\eqref{eq:M0ss} is the stronger assumption;
the consistency check above limits the departure from additivity to
$\Delta_s^{\rm eff}-\Delta_s\simeq+4\pm55$~MeV, compatible with zero within
present uncertainties. A nonlinear increment, should one be established, would
shift the two triplets rigidly without altering their internal spacings, so
the relational predictions---the fixed spacings and the near-degeneracy of the
upper $S$-class pair---are robust against this assumption.

This robustness has a physical basis worth making explicit, since
Eq.~\eqref{eq:M0ss} sets the absolute scale. The increment
$\Delta_s=127^{+6}_{-5}$~MeV is comparable to the constituent strange--light
mass difference $m_s-m_q$---the part of the baseline shift that is additive by
construction---while the spin-dependent contributions, which could in
principle be nonlinear, do not enter $M_0$: they are carried separately by the
exchange couplings $J^{ss}$ and are removed by the spin-averaging that defines
the baseline. Strange-mass additivity at this level is moreover an established
regularity of the light-baryon sector from which the couplings are drawn;
its transfer to the doubly-strange pentaquark baseline is what we adopt as the
additivity ansatz, and serves as a plausibility argument rather than a proof.
The decuplet $\Delta(1232)$, $\Sigma^*(1385)$, $\Xi^*(1530)$, $\Omega(1672)$ is
equally spaced to within $\sim10$~MeV ($153$, $145$, and $142$~MeV per added
strange quark), so each strange substitution shifts the mass by a nearly
constant amount; the mild downward curvature ($153\!\to\!142$~MeV) indicates
that any nonadditivity is small and, if anything, slightly sub-additive,
which would lower $M_0^{(ss)}$ by $\sim10$--$20$~MeV rather than raise it.

A second potential source of nonlinearity is the core--cloud binding, which
changes as the heavier $ssq$ cloud becomes more compact. It is bounded by the
same logic: the singly-strange $\Delta_s$ already absorbs the binding response
to the first strange substitution, so any deviation from $2\Delta_s$ is the
difference between the first and second response---second order in the
modest change of the cloud's reduced mass---and is expected at the
$\lesssim\!10$--$20$~MeV level. Alternative extrapolations confirm this
bracket: a decuplet-calibrated increment ($\sim145$~MeV per quark) raises
$M_0^{(ss)}$ by $\sim36$~MeV, a constituent estimate $2(m_s-m_q)$ with
$m_s-m_q\approx130$~MeV reproduces $2\Delta_s$ to within $\sim10$~MeV, and a
$\pm10\%$ nonadditivity in the second-quark increment moves the scale by only
$\sim\pm13$~MeV. Each amounts to a rigid common shift of at most a few tens of
MeV, leaving the predictions above unchanged; in this sense the splittings
introduce no free parameter of their own, following without adjustment from
the $J^{qq}$ couplings and the scaling factor fixed by the measured $P_c$ and
$P_{cs}$ spectra, while the absolute scale rests in addition on the
single additive ansatz.

To make the impact of a possible departure from additivity explicit, we write
\begin{equation}
  M_0^{(ss)} = M_0 + 2\Delta_s + \delta ,
  \label{eq:nonadd}
\end{equation}
with $\delta$ measuring the nonadditivity of the strange-mass increment. Because
$\delta$ is a property of the baseline extrapolation and not of the light-cloud
couplings, it is common to both production classes and enters every state
additively: each predicted mass shifts by exactly $\delta$, while the internal
splittings, the $125$~MeV gap between the two classes, the $\sim70$~MeV
$S$-class ground-to-doublet separation, the $\sim4$~MeV doublet splitting, and
the level ordering are all independent of $\delta$. The three extrapolations
above bracket its size: the decuplet calibration gives $\delta\simeq+36$~MeV,
the constituent estimate $\delta\simeq0$, and a $\pm10\%$ nonadditivity
$\delta\simeq\pm13$~MeV, while the mild sub-additivity of the light-baryon
decuplet favors a small negative value, so a conservative window is
$-20\lesssim\delta\lesssim+36$~MeV. Table~\ref{tab:nonadd} lists the resulting
masses across this window. Even at the extremes the spectrum moves rigidly by
at most a few tens of MeV---well inside the $\sim\pm50$~MeV additivity entry of
Table~\ref{tab:budget}---so moderate nonadditivity rescales the absolute
predictions without disturbing any of their relational or qualitative content.

\begin{table}[!h]
\caption{Predicted masses (MeV) under a nonadditivity offset $\delta$ in
$M_0^{(ss)}=M_0+2\Delta_s+\delta$ [Eq.~\eqref{eq:nonadd}], for representative
$\delta$ spanning the conservative window $-20\lesssim\delta\lesssim+36$~MeV.
Every state shifts rigidly by $\delta$; the internal splittings and the level
ordering are unchanged. Shown are the lowest $S$-class state, the
near-degenerate $S$-class doublet, and the lowest $A$-class state.}
\label{tab:nonadd}
\begin{ruledtabular}
\begin{tabular}{lcccc}
State & $\delta=-20$ & $\delta=0$ & $\delta=+20$ & $\delta=+36$ \\
\hline
$P_{css}(\tfrac12^-)$        & 4583 & 4603 & 4623 & 4639 \\
$P_{css}'(\tfrac12^-)$       & 4655 & 4675 & 4695 & 4711 \\
$P_{css}''(\tfrac32^-)$      & 4659 & 4679 & 4699 & 4715 \\
$\tilde P_{css}(\tfrac12^-)$ & 4461 & 4481 & 4501 & 4517 \\
\end{tabular}
\end{ruledtabular}
\end{table}

A realistic error budget combines the coupling-fit errors of
Table~\ref{tab:spectrum} with these systematics, collected in
Table~\ref{tab:budget}; they separate cleanly into effects on the internal
splittings and on the common absolute scale. The chromomagnetic
ratio is the only one that acts on the splittings: scanning
$\kappa^{qs}/\kappa^{qq}$ over $0.5$--$0.7$ shifts the individual $\tfrac12^-$
states by up to $\mp8$--$9$~MeV---the two members of a pair moving oppositely
as the splitting widens---the $S$-class $\tfrac32^-$ by $\pm5$~MeV, and the
$A$-class $\tfrac32^-$ by only $\pm1$~MeV, the baseline being held fixed; the
explicit shifts are listed in Table~\ref{tab:kappa}. The remaining systematics
act on the scale alone. The additive increment contributes $\sim\pm50$~MeV, as
discussed above, but cancels in the splittings. With $S_{c\bar c}=0$ the
heavy-core hyperfine interaction vanishes at leading order and enters only
through $S_{c\bar c}=1$ admixture, suppressed by $1/m_c^2$ and the core
spin-splitting; we assign it $\lesssim10$~MeV, common to a triplet. Mixing of
the $\mathbf 8 \otimes \mathbf 8$ cloud with color-singlet ($\mathbf 1 \otimes \mathbf1$) hadronic configurations
is the least controlled effect; assumed negligible for the narrow states in
Ref.~\cite{Germani:2024miu}, its size can nonetheless be estimated
systematically in second-order perturbation theory. Coupling to the nearby
$\mathbf 1\otimes \mathbf 1$ two-hadron channels $n$ shifts the mass by
\begin{equation}
  \delta M \simeq \sum_n \frac{|g_n|^2}{M-E_n},
  \label{eq:mixshift}
\end{equation}
with $g_n$ the fall-apart coupling to channel $n$ and $M-E_n$ the distance to
its threshold. Two inputs fix the scale without a full coupled-channel solve.
First, the same coupling generates the fall-apart width,
$\Gamma\simeq2\pi\sum_n|g_n|^2\rho_n$ with $\rho_n$ the phase-space density;
the observed $P_c$ and $P_{cs}$ have $\Gamma\sim10$--$40$~MeV, which bounds
$|g_n|^2\rho_n$ to at most a few MeV per channel. Second, the denominators are
read directly from Table~\ref{tab:thresholds}: the relevant open- and
near-threshold channels lie $\sim20$--$200$~MeV away. Relating the real shift
to the width through the usual dispersion relation, $\delta M$ is of order
$(\Gamma/2\pi)$ times a principal-value factor of order unity summed over the
handful of channels, giving $\delta M\sim10$--$20$~MeV---of either sign,
according as the nearest threshold lies above or below the state, and largest
for the near-threshold members, where the smallest denominator in
Eq.~\eqref{eq:mixshift} dominates. We stress that this remains a parametric,
order-of-magnitude estimate rather than a derivation from within the present
scheme: the model does not itself specify $g_n$, which we have bounded
externally through the observed widths, and a firm value would require solving
the coupled-channel problem explicitly. We adopt the resulting $\sim20$~MeV only
to gauge the size of the associated systematic. This is also the sense in which the ``no fitted parameters''
description must be qualified: color-singlet mixing is a real effect of the
underlying dynamics whose magnitude the scheme does not control, so the absence
of fitted parameters reflects the economy of the construction rather than
complete control of the dynamics. Added in quadrature, these give a realistic uncertainty
of $\sim\pm55$~MeV on the absolute masses---dominated by additivity and
color-singlet mixing---and only $\sim\pm13$~MeV on the internal splittings. The
residual $\sim4$~MeV gap of the upper $S$-class pair lies within these
systematics, reinforcing that the experimentally robust feature is the
near-degeneracy itself rather than the resolved ordering.

\begin{table}[!h]
\caption{Realistic error budget for the predicted masses (MeV), separating
effects on the internal splittings from those on the common (absolute) mass
scale; entries are added in quadrature.}
\label{tab:budget}
\begin{ruledtabular}
\begin{tabular}{lcc}
Source & Splittings & Abs.\ scale \\
\hline
$J^{qq}$ coupling fit & $\pm3$ & -- \\
$\Delta_s$ extraction & -- & $\pm12$ \\
$\kappa^{qs}/\kappa^{qq}\in[0.5,0.7]$ & $\pm8$ & -- \\
additivity, $M_0^{(ss)}=M_0+2\Delta_s$ & -- & $\pm50$ \\
heavy-core hyperfine ($S_{c\bar c}=0$) & $\lesssim3$ & $\lesssim10$ \\
$8\otimes8$--$1\otimes1$ mixing & $\sim10$ & $\sim20$ \\
\hline
total (quadrature) & $\pm13$ & $\pm56$ \\
\end{tabular}
\end{ruledtabular}
\end{table}

\begin{table}[!h]
\caption{Mass shifts (MeV) of the predicted states when the chromomagnetic
ratio $\kappa^{qs}/\kappa^{qq}$ is moved from its central value $0.6$ to the
edges of the conservative range $0.5$--$0.7$. The baseline is held fixed, so
these measure the sensitivity of the splittings alone.}
\label{tab:kappa}
\begin{ruledtabular}
\begin{tabular}{lccc}
State & $J^P$ & $\kappa=0.5$ & $\kappa=0.7$ \\
\hline
$P_{css}$        & $\tfrac12^-$ & $+8$ & $-8$ \\
$P_{css}'$       & $\tfrac12^-$ & $-8$ & $+8$ \\
$P_{css}''$      & $\tfrac32^-$ & $-5$ & $+5$ \\
$\tilde P_{css}$  & $\tfrac12^-$ & $+8$ & $-9$ \\
$\tilde P_{css}'$ & $\tfrac32^-$ & $-1$ & $0$ \\
$\tilde P_{css}''$& $\tfrac12^-$ & $-8$ & $+9$ \\
\end{tabular}
\end{ruledtabular}
\end{table}

When considered together, these comparisons provide a coherent interpretation. The
baryo-charmonium scheme places the doubly-strange states in the same
$4.4$--$4.7$~GeV window favored by the molecular and sum-rule analyses, but
reaches it without introducing new fitted parameters and with a markedly
different internal organization: where the molecular poles are individually
anchored to nearby two-hadron thresholds, ours form two production-defined
triplets with fixed internal spacings. The level-ordering inversion of the $S$ class and the near-degeneracy of its upper pair are, in this light, clean
predictions that carry no adjustable parameter of their own---they depend only
on the light-cloud couplings, not on the additive scale---and that an
amplitude analysis of the $J/\psi\,\Xi$ distribution could confront.

A final assumption warrants comment, since it is the one most directly in
tension with the molecular interpretations: that the narrow states are
dominantly the compact $(c\bar c)_8(qqq)_8$ configuration, with color-singlet
($\mathbf 1\otimes \mathbf1$) admixtures small. Several considerations bear on it, and they do
not all point the same way. The nearby
open-charm channels---$\Xi_c\bar D_s$, $\Xi_c'\bar D_s$, $\Xi_c\bar D_s^*$, and
$\Omega_c\bar D^{(*)}$---are reached only by dissociating the heavy pair, with
one $c$ entering a charmed baryon and the $\bar c$ a charmed-strange meson; for
heavy quarks this rearrangement is suppressed, as the tightly bound octet
$(c\bar c)$ core must be separated over a hadronic distance. The compact state
and these loosely bound meson--baryon configurations also differ greatly in
spatial extent, so their overlap---and with it the fall-apart coupling---is
small. Both effects, moreover, point in a favorable direction as strangeness is
added. The heavy-quark rearrangement that suppresses the fall-apart coupling is
controlled by the compactness of the $(c\bar c)$ core, which is flavor-blind,
so the suppression is no weaker for $c\bar c ssq$ than for the lighter sectors;
and the $ssq$ cloud, built from heavier constituents, is more spatially compact
than the $uud$ or $uds$ clouds, reducing still further its overlap with the
extended molecular configurations. On these counts the $\mathbf 8 \otimes \mathbf8$ dominance
is, if anything, better motivated in the doubly-strange sector than in the
cases where it is already supported by data. Empirically the same suppression
already operates one and two steps down
in strangeness: the observed $P_c$ and $P_{cs}$ that fix the couplings are
narrow ($\Gamma\sim$ tens of MeV), which directly bounds their $\mathbf 1\otimes \mathbf 1$
content. One sector-specific effect cuts the other way, however, and is the
genuine source of risk: the doubly-strange states lie closer to---and the
$S$-class members above several of---the open-charm thresholds, and
near-threshold states are the most susceptible to $\mathbf 1\otimes \mathbf1$ continuum mixing.
This is precisely the feature the molecular analyses exploit in treating these
states as dominantly $\mathbf 1\otimes \mathbf1$, and it is why our mixing systematic is
assigned channel by channel and is largest for the near-threshold members
(Table~\ref{tab:budget}). We therefore present the $\mathbf 8\otimes \mathbf8$ dominance not as
established but as a working hypothesis whose sector-specific validity is itself
the central physical question; the distinctive pattern predicted here---two
fixed-spacing production triplets with a near-degenerate upper pair---is offered
as the experimental discriminator that would confirm or refute it. On balance we
expect---though we do not prove---that mixing remains a
perturbation here as well, and have entered its heuristically estimated effect,
$\sim20$~MeV and channel-dependent, in the scale budget of
Table~\ref{tab:budget}; a quantitative value lies beyond the present scheme,
which does not contain the $\mathbf 8\otimes \mathbf8$--$\mathbf 1\otimes \mathbf1$ coupling, and would require
an explicit coupled-channel calculation.

The consequences for the predicted ordering are modest but worth stating. The
open-charm thresholds carry definite quantum numbers---the pseudoscalar
channels $\Xi_c^{(\prime)}\bar D_s$ and $\Omega_c\bar D$ couple to $\tfrac12^-$,
$\Xi_c^*\bar D_s$ to $\tfrac32^-$, and the vector channels $\Xi_c\bar D_s^*$ and
$\Omega_c\bar D^*$ to both---so mixing can displace the $\tfrac12^-$ and
$\tfrac32^-$ members unequally. The mechanism producing the inversion,
the large light-cloud ratio $|J_A^{qq}/J_S^{qq}|$ that drives $\Delta E_{3/2}$
above $\Delta E_{1/2}$, is a short-distance effect and is not removed by
long-range threshold coupling. The numerical margin is small, however
($\sim4$~MeV for the upper $S$-class pair), so $J^P$-dependent mixing of this
size could blur or reorder that pair; the mixing-insensitive content of the
prediction is accordingly the appearance of two production-defined triplets
with the $\tfrac32^-$ driven to the top of a near-degenerate upper pair, rather
than the resolved $4$~MeV ordering within it. Threshold effects, if present,
would show up as cusps in the $J/\psi\,\Xi$ line shape near the open-charm
thresholds rather than as large mass displacements, so long as mixing stays
perturbative.

The two pictures are, moreover, experimentally separable. A molecular state is
pinned to its dominant meson--baryon threshold and shifts if the binding
changes, so the molecular analyses predict a sequence of threshold-tracking
poles of mixed parity, including $\tfrac52^-$ and positive-parity members. A
baryo-charmonium triplet instead keeps fixed internal spacings set by the
light-quark couplings, independent of where the multiplet sits, and---under the
minimal $S$-wave, $S_{c\bar c}=0$ assumption---contains only negative-parity
$\tfrac12^-$ and $\tfrac32^-$ states. Counting the states in the
$J/\psi\,\Xi$ spectrum, determining their parities, and measuring their
relative spacings would therefore discriminate directly between a
threshold-driven and a compact internal structure; the robust $S$-class
discriminator the present scheme provides is the near-degeneracy of the upper
pair---a doublet whose $\tfrac12^-$--$\tfrac32^-$ spacing has collapsed from the
$\sim40$~MeV of the lighter sectors to only $\sim4$~MeV---rather than the
resolved ordering within it.
We stress that the upper $S$-class pair is split by only
$\sim4$~MeV, so resolving the two states and assigning their spins lies beyond
present $J/\psi\,\Xi$ mass resolution; the immediately accessible signatures
are therefore the near-degenerate doublet near $4.68$~GeV and its $\sim70$~MeV
separation from the lower $\tfrac12^-$ state at $4.60$~GeV, with the strict
$\tfrac32^-$-on-top ordering becoming a direct experimental test only once
sub-$10$~MeV resolution and a spin analysis are in hand.

% ====================================================================
\section{Production and decay channels}
\label{sec:prod}
% ====================================================================
The two production classes correspond to two distinct hadronic environments,
inherited from the color-flavor symmetry of the spectator light system that
survives into the final state~\cite{Germani:2024miu}. The $S$ class is produced in
association with a light meson---as in $\Lambda_b^0\!\to\!P_cK^-$ and
$\Xi_b^-\!\to\!P_{cs}K^-$, where the ``good'' light diquark retains its
symmetric color-flavor configuration---whereas the $A$ class is produced in
association with an antibaryon, as in $B^-\!\to\!\tilde P_{cs}\bar p$.

For the doubly-strange $S$-class states the relevant processes are already in
hand. The CMS and LHCb Collaborations have observed the $S=-2$ decays
$\Lambda_b^0\!\to\!J/\psi\,\Xi^-K^+$ and
$\Xi_b^0\!\to\!J/\psi\,\Xi^-\pi^+$~\cite{CMS:2024vnm,LHCb:2025lhk}, whose $J/\psi\,\Xi^-$
subsystem carries precisely the $c\bar c ssd$ content of a $P_{css}^-$. A
$P_{css}$ would therefore show up as a peak in the $J/\psi\,\Xi$
invariant-mass distribution of these decays---the same channel emphasized in
the molecular analyses~\cite{Roca:2024nsi,Clymton:2025zer}---and decays of heavier doubly-strange
$b$ baryons, such as the $\Omega_b^-$, could feed the sector as statistics
grow. The $A$-class $\tilde P_{css}$ instead calls for antibaryon-associated
production, the analog of $B^-\!\to\!\tilde P_{cs}\bar p$; lacking such data,
the $S$ class is the immediate experimental target.

\onecolumngrid
\begin{table}[!h]
\centering
\caption{Excess energy $\delta_X=M-M(X)$ (MeV) of each predicted $P_{css}$
state above the relevant hidden-charm and open-charm two-hadron thresholds
$X$ (threshold masses, MeV, in parentheses). Negative entries denote
kinematically closed channels. The $A$-class states clear only the lowest
open-charm threshold $\Xi_c\bar D_s$, whereas the $S$-class states lie above
most of the open-charm ladder---consistent with the expected
narrow-to-broad trend.}
\label{tab:thresholds}
\begin{ruledtabular}
\begin{tabular}{lccccccc}
State & $J^P$ & Mass & $J/\psi\,\Xi$ & $\Xi_c\bar D_s$ & $\Xi_c'\bar D_s$ & $\Xi_c\bar D_s^*$ & $\Xi_c^*\bar D_s$ \\
      &       &      & (4415)        & (4437)          & (4547)           & (4581)            & (4614) \\
\hline
$P_{css}$        & $\tfrac12^-$ & 4603 & 188 & 166 & 56   & 22    & $-11$  \\
$P_{css}'$       & $\tfrac12^-$ & 4675 & 260 & 238 & 128  & 94    & 61     \\
$P_{css}''$      & $\tfrac32^-$ & 4679 & 264 & 242 & 132  & 98    & 65     \\[3pt]
$\tilde P_{css}$  & $\tfrac12^-$ & 4481 & 66  & 44  & $-66$  & $-100$ & $-133$ \\
$\tilde P_{css}'$ & $\tfrac32^-$ & 4534 & 119 & 97  & $-13$  & $-47$  & $-80$  \\
$\tilde P_{css}''$& $\tfrac12^-$ & 4547 & 132 & 110 & 0    & $-34$  & $-67$  \\
\end{tabular}
\end{ruledtabular}
\end{table}
\twocolumngrid

The expected decay pattern follows from the predicted masses
(Fig.~\ref{fig:levels}). Every state lies above the hidden-charm $\eta_c\Xi$
($\sim4300$~MeV) and $J/\psi\,\Xi$ ($\sim4415$~MeV) thresholds and above the
open-charm $\Xi_c\bar D_s$ threshold ($\sim4437$~MeV), so all three channels
are open throughout, with $J/\psi\,\Xi$ the natural discovery mode. The two
lowest states sit only a few tens of MeV above $\Xi_c\bar D_s$ and have little
open-charm phase space, so they are expected to be narrow; the heavier members
lie among the $\Xi_c'\bar D_s$, $\Xi_c\bar D_s^*$, and $\Xi_c^*\bar D_s$
thresholds, which open further open-charm modes and would on this basis be
expected to be broader. This
ordering---narrow states at the bottom of each triplet, broader ones above---is
an additional qualitative handle for experimental identification.

Table~\ref{tab:thresholds} makes this pattern quantitative. All six states lie
well above the $J/\psi\,\Xi$ discovery channel, by
$\delta_{J/\psi\Xi}=66$--$264$~MeV, so it serves as the universal search mode;
the open-charm channels, by contrast, switch on progressively. The three
$A$-class states clear only the lowest open-charm threshold $\Xi_c\bar D_s$:
the ground state $\tilde P_{css}(\tfrac12^-)$ sits just $44$~MeV above it, and
the two heavier members remain at or below the next threshold
$\Xi_c'\bar D_s$ ($\delta_{\Xi_c'\bar D_s}\le0$), the highest lying essentially
on it. With a single open-charm mode and little phase space, the $A$-class
triplet is expected to be narrow. The $S$-class states lie far higher up the ladder:
$P_{css}(\tfrac12^-)$ has three open-charm channels accessible
($\Xi_c\bar D_s$, $\Xi_c'\bar D_s$, $\Xi_c\bar D_s^*$), while the
near-degenerate pair $P_{css}'(\tfrac12^-)$ and $P_{css}''(\tfrac32^-)$ clears
all four, $\Xi_c^*\bar D_s$ included. The number of open open-charm channels
therefore rises from one for the $A$ class to three or four for the $S$ class,
sharpening the qualitative narrow-to-broad expectation into a concrete one:
on this counting one would expect a narrower $J/\psi\,\Xi$ peak near
$4.48$~GeV from the $A$-class ground state, accompanied by broader structures
in the $4.60$--$4.68$~GeV region where the
$S$-class triplet sits atop the open-charm ladder. We emphasize that this
narrow-to-broad expectation rests on channel counting and available phase
space alone: no partial widths are computed here, since the fall-apart
couplings depend on the $\mathbf 8\otimes \mathbf 8$--$\mathbf  1\otimes \mathbf 1$ overlap that the scheme does
not fix. The pattern should therefore be read as an ordering of relative
widths within and between the two triplets rather than as a quantitative
prediction, and a dynamical width calculation is left to future work.

% ====================================================================
\section{Summary}
\label{sec:summary}
% ====================================================================
We have extended the baryo-charmonium scheme to the doubly-strange
hidden-charm sector without introducing any new fitted parameter. The
splittings carry no adjustable parameter of their own, being inherited without
adjustment from the observed $P_c$ and $P_{cs}$ states through the light-quark
$J^{qq}$ couplings and the assumed chromomagnetic scaling; the absolute scale,
by contrast, rests in addition on a single additive strange-mass increment, which we
have been able to test only against another
theoretical estimate and which remains the load-bearing assumption of the
work. The resulting spectrum comprises two
negative-parity triplets in the $4.5$--$4.7$~GeV range. Its robust qualitative
feature is that the upper two kaon-associated states collapse into a
near-degenerate pair---their spacing falling from the $\sim40$~MeV of the
lighter sectors to only $\sim4$~MeV, a consequence of the chromomagnetic
suppression of the symmetric $ss$ coupling [Eq.~\eqref{eq:crossover}] and
stable against the fitted coupling uncertainties.
Within that pair the level ordering formally inverts, placing the $\tfrac32^-$
on top. Its significance is set not by the $\sim4$~MeV margin against the
$\sim13$~MeV per-level systematic---the relevant systematics being common-mode
and cancelling in the splitting difference that controls the sign---but by the
direct propagation into that difference, which leaves the inversion realized in
$96\%$ of samples ($\approx1.8\sigma$). We therefore present it as a
$\sim2\sigma$ preference and a qualitative tendency rather than a resolved
prediction, the doublet itself being the firm result. Inverting the construction about an independent QCD sum-rule
mass returns an effective strange-mass increment numerically consistent with the
singly-strange value. We emphasize, however, that this comparison is only a
weak check: the $\pm55$~MeV uncertainty admits moderate departures from
additivity, and the test in the pentaquark sector is against another
theoretical estimate, not data. Its physical content---strange-mass
additivity---is, on the other hand, a measured regularity of the light-baryon
decuplet, so what remains a working assumption rather than an established
result is specifically its transfer to the doubly-strange pentaquark baseline.
A single measured doubly-strange mass would fix that baseline and test the
assumption directly. The absolute mass scale---though bracketed by the
systematics of Sec.~\ref{sec:check}---should meanwhile be regarded as the least
secure part of the prediction; the predicted masses are nonetheless consistent
with molecular coupled-channel calculations. Because the
upper $S$-class pair is split by only $\sim4$~MeV---below present
$J/\psi\,\Xi$ resolution---the experimentally robust content is the
near-degenerate doublet near $4.68$~GeV and its $\sim70$~MeV separation from
the lower $\tfrac12^-$ state at $4.60$~GeV, the strict $\tfrac32^-$-on-top
ordering becoming a direct test only with sub-$10$~MeV resolution and a spin
analysis. These states should be accessible at LHCb in the $J/\psi\,\Xi$ and
$\Xi_c\bar D_s$ final states now opening up through $S=-2$ $b$-baryon decays.

\begin{acknowledgments}
This work is supported by Scientific Research Projects Coordination Unit of Ondokuz Mayis University with project
BAP05-2025-5384.
\end{acknowledgments}

\appendix
\section{Doubly-strange splittings and the inversion condition}
\label{app:deriv}
The three pair couplings entering Eqs.~\eqref{eq:dE12}--\eqref{eq:dE32} are
read directly off the color-flavor tensors~\eqref{eq:Sssq}--\eqref{eq:Assq}.
In the $S$ class $(J_a,J_b,J_c)=(J_S^{ss},J_A^{qs},J_A^{qs})$; since two of the
three couplings are equal, the square root in Eq.~\eqref{eq:dE12} collapses and
\begin{align}
  \Delta E_{1/2}^{S}&=\big|\,J_S^{ss}-J_A^{qs}\,\big|,\label{eq:dE12S}\\
  \Delta E_{3/2}^{S}&=-\big(J_S^{ss}+2J_A^{qs}\big).\label{eq:dE32S}
\end{align}
In the $A$ class $(J_a,J_b,J_c)=(J_S^{qs},J_A^{ss},J_S^{qs})$ and the overall
signs reverse, giving
\begin{align}
  \Delta E_{1/2}^{A}&=\big|\,J_S^{qs}-J_A^{ss}\,\big|,\label{eq:dE12A}\\
  \Delta E_{3/2}^{A}&=2J_S^{qs}+J_A^{ss}.\label{eq:dE32A}
\end{align}
Inserting the couplings of Table~\ref{tab:couplings} reproduces the splittings
quoted in Sec.~\ref{sec:spectrum}.

The level ordering follows from the sign of
$\Delta E_{3/2}-\Delta E_{1/2}$. With $J_S^{ss}>0$ and $J_A^{qs}<0$,
Eqs.~\eqref{eq:dE12S}--\eqref{eq:dE32S} give the simple difference
\begin{equation}
  \Delta E_{3/2}^{S}-\Delta E_{1/2}^{S}=\big|J_A^{qs}\big|-2\,J_S^{ss},
\end{equation}
which is positive---the inversion---precisely when $|J_A^{qs}|>2J_S^{ss}$.
Using $|J_A^{qs}|=(\kappa^{qs}/\kappa^{qq})|J_A^{qq}|$ and
$J_S^{ss}=(\kappa^{qs}/\kappa^{qq})^2 J_S^{qq}$ this reduces to the compact
condition of Eq.~\eqref{eq:crossover}. The inversion is thus driven by the
large ratio $|J_A^{qq}/J_S^{qq}|$ of the light cloud, only partially offset by
the chromomagnetic suppression of the strange couplings, and is specific to
the doubly-strange $S$ class: the same difference evaluated for the $A$ class,
or for the singly-strange and non-strange triplets, stays negative.

\bibliographystyle{apsrev4-2}
\bibliography{Pcss_paper}

@article{Aaij:2015tga,
    author = "Aaij, Roel and others",
    collaboration = "LHCb",
    title = "{Observation of $J/\psi p$ Resonances Consistent with Pentaquark States in $\Lambda_b^0 \to J/\psi K^- p$ Decays}",
    eprint = "1507.03414",
    archivePrefix = "arXiv",
    primaryClass = "hep-ex",
    reportNumber = "CERN-PH-EP-2015-153, LHCB-PAPER-2015-029",
    doi = "10.1103/PhysRevLett.115.072001",
    journal = "Phys. Rev. Lett.",
    volume = "115",
    pages = "072001",
    year = "2015"
}

@article{Aaij:2019vzc,
    author = "Aaij, Roel and others",
    collaboration = "LHCb",
    title = "{Observation of a narrow pentaquark state, $P_c(4312)^+$, and of two-peak structure of the $P_c(4450)^+$}",
    eprint = "1904.03947",
    archivePrefix = "arXiv",
    primaryClass = "hep-ex",
    reportNumber = "LHCb-PAPER-2019-014 CERN-EP-2019-058",
    doi = "10.1103/PhysRevLett.122.222001",
    journal = "Phys. Rev. Lett.",
    volume = "122",
    number = "22",
    pages = "222001",
    year = "2019"
}

@article{LHCb:2020jpq,
    author = "Aaij, Roel and others",
    collaboration = "LHCb",
    title = "{Evidence of a $J/\psi\Lambda$ structure and observation of excited $\Xi^-$ states in the $\Xi^-_b \to J/\psi\Lambda K^-$ decay}",
    eprint = "2012.10380",
    archivePrefix = "arXiv",
    primaryClass = "hep-ex",
    reportNumber = "LHCb-PAPER-2020-039, CERN-EP-2020-233",
    doi = "10.1016/j.scib.2021.02.030",
    journal = "Sci. Bull.",
    volume = "66",
    pages = "1278--1287",
    year = "2021"
}

@article{LHCb:2022ogu,
    author = "Aaij, R. and others",
    collaboration = "LHCb",
    title = "{Observation of a J/{\ensuremath{\psi}}{\ensuremath{\Lambda}} Resonance Consistent with a Strange Pentaquark Candidate in B-{\textrightarrow}J/{\ensuremath{\psi}}{\ensuremath{\Lambda}}p{\textasciimacron} Decays}",
    eprint = "2210.10346",
    archivePrefix = "arXiv",
    primaryClass = "hep-ex",
    reportNumber = "CERN-EP-2022-198, LHCb-PAPER-2022-031",
    doi = "10.1103/PhysRevLett.131.031901",
    journal = "Phys. Rev. Lett.",
    volume = "131",
    number = "3",
    pages = "031901",
    year = "2023"
}

@article{Belle:2025pey,
    author = "Adachi, I. and others",
    collaboration = "Belle, Belle-II",
    title = "{Search for Pcs(4459) and Pcs(4338) in Upsilon(1S,2S) inclusive decays at Belle}",
    eprint = "2502.09951",
    archivePrefix = "arXiv",
    primaryClass = "hep-ex",
    reportNumber = "Belle II Preprint 2025-002, KEK Preprint 2024-50",
    doi = "10.1103/pf8m-6j69",
    journal = "Phys. Rev. Lett.",
    volume = "135",
    number = "4",
    pages = "041901",
    year = "2025"
}

@article{Maiani:2023nwj,
    author = "Maiani, Luciano and Polosa, Antonio D. and Riquer, Veronica",
    title = "{The pentaquark spectrum from Fermi statistics}",
    eprint = "2303.04056",
    archivePrefix = "arXiv",
    primaryClass = "hep-ph",
    doi = "10.1140/epjc/s10052-023-11492-0",
    journal = "Eur. Phys. J. C",
    volume = "83",
    number = "5",
    pages = "378",
    year = "2023"
}

@article{Esposito:2016noz,
    author = "Esposito, A. and Pilloni, A. and Polosa, A. D.",
    title = "{Multiquark Resonances}",
    eprint = "1611.07920",
    archivePrefix = "arXiv",
    primaryClass = "hep-ph",
    reportNumber = "JLAB-THY-16-2301",
    doi = "10.1016/j.physrep.2016.11.002",
    journal = "Phys. Rept.",
    volume = "668",
    pages = "1--97",
    year = "2017"
}

@article{Meng:2022ozq,
    author = "Meng, Lu and Wang, Bo and Wang, Guang-Juan and Zhu, Shi-Lin",
    title = "{Chiral perturbation theory for heavy hadrons and chiral effective field theory for heavy hadronic molecules}",
    eprint = "2204.08716",
    archivePrefix = "arXiv",
    primaryClass = "hep-ph",
    doi = "10.1016/j.physrep.2023.04.003",
    journal = "Phys. Rept.",
    volume = "1019",
    pages = "1--149",
    year = "2023"
}

@article{Germani:2024miu,
    author = "Germani, Davide and Niliani, Farhad and Polosa, Antonio D.",
    title = "{A model of pentaquarks}",
    eprint = "2403.04068",
    archivePrefix = "arXiv",
    primaryClass = "hep-ph",
    doi = "10.1140/epjc/s10052-024-13103-y",
    journal = "Eur. Phys. J. C",
    volume = "84",
    number = "7",
    pages = "755",
    year = "2024"
}

@article{Anisovich:2015zqa,
    author = "Anisovich, V. V. and Matveev, M. A. and Nyiri, J. and Sarantsev, A. V. and Semenova, A. N.",
    title = "{Nonstrange and strange pentaquarks with hidden charm}",
    eprint = "1509.04898",
    archivePrefix = "arXiv",
    primaryClass = "hep-ph",
    doi = "10.1142/S0217751X15501900",
    journal = "Int. J. Mod. Phys. A",
    volume = "30",
    number = "32",
    pages = "1550190",
    year = "2015"
}

@article{Meng:2019fan,
    author = "Meng, Qi and Hiyama, Emiko and Can, Kadir Utku and Gubler, Philipp and Oka, Makoto and Hosaka, Atsushi and Zong, Hongshi",
    title = "{Compact $sssc\bar{c}$ pentaquark states predicted by a quark model}",
    eprint = "1907.00144",
    archivePrefix = "arXiv",
    primaryClass = "nucl-th",
    doi = "10.1016/j.physletb.2019.135028",
    journal = "Phys. Lett. B",
    volume = "798",
    pages = "135028",
    year = "2019"
}

@article{Wang:2020bjt,
    author = "Wang, Fu-Lai and Chen, Rui and Liu, Xiang",
    title = "{Prediction of hidden-charm pentaquarks with double strangeness}",
    eprint = "2011.14296",
    archivePrefix = "arXiv",
    primaryClass = "hep-ph",
    doi = "10.1103/PhysRevD.103.034014",
    journal = "Phys. Rev. D",
    volume = "103",
    number = "3",
    pages = "034014",
    year = "2021"
}

@article{Ferretti:2020ewe,
    author = "Ferretti, J. and Santopinto, E.",
    title = "{Hidden-charm and bottom tetra- and pentaquarks with strangeness in the hadro-quarkonium and compact tetraquark models}",
    eprint = "2001.01067",
    archivePrefix = "arXiv",
    primaryClass = "hep-ph",
    doi = "10.1007/JHEP04(2020)119",
    journal = "JHEP",
    volume = "04",
    pages = "119",
    year = "2020"
}

@article{Ferretti:2021zis,
    author = "Ferretti, J. and Santopinto, E.",
    title = "{The new $P_{\rm cs}(4459)$, $Z_{\rm cs}(3985)$, $Z_{\rm cs}(4000)$ and $Z_{\rm cs}(4220)$ and the possible emergence of flavor pentaquark octets and tetraquark nonets}",
    eprint = "2111.08650",
    archivePrefix = "arXiv",
    primaryClass = "hep-ph",
    doi = "10.1016/j.scib.2022.04.010",
    journal = "Sci. Bull.",
    volume = "67",
    pages = "1209",
    year = "2022"
}

@article{Wang:2021hql,
    author = "Wang, Fu-Lai and Yang, Xin-Dian and Chen, Rui and Liu, Xiang",
    title = "{Hidden-charm pentaquarks with triple strangeness due to the $\Omega_{c}^{(*)}\bar{D}_s^{(*)}$ interactions}",
    eprint = "2101.11200",
    archivePrefix = "arXiv",
    primaryClass = "hep-ph",
    doi = "10.1103/PhysRevD.103.054025",
    journal = "Phys. Rev. D",
    volume = "103",
    number = "5",
    pages = "054025",
    year = "2021"
}

@article{Roca:2024nsi,
    author = "Roca, L. and Song, J. and Oset, E.",
    title = "{Molecular pentaquarks with hidden charm and double strangeness}",
    eprint = "2403.08732",
    archivePrefix = "arXiv",
    primaryClass = "hep-ph",
    doi = "10.1103/PhysRevD.109.094005",
    journal = "Phys. Rev. D",
    volume = "109",
    number = "9",
    pages = "094005",
    year = "2024"
}

@article{Clymton:2025zer,
    author = "Clymton, Samson and Kim, Hyun-Chul and Mart, Terry",
    title = "{Double-strangeness hidden-charm pentaquarks}",
    eprint = "2506.23587",
    archivePrefix = "arXiv",
    primaryClass = "hep-ph",
    reportNumber = "INHA-NTG-04/2025",
    doi = "10.1103/gy7h-9dt5",
    journal = "Phys. Rev. D",
    volume = "112",
    number = "3",
    pages = "034015",
    year = "2025"
}

@article{Roca:2025zyi,
    author = "Roca, L. and Song, J. and Oset, E.",
    title = "{Study of hidden-charm, doubly-strange pentaquarks in $\Lambda _b\rightarrow J/\psi \Xi ^- K^+$ and $\Xi _b\rightarrow J/\psi \Xi ^- \pi ^+$}",
    eprint = "2509.19840",
    archivePrefix = "arXiv",
    primaryClass = "hep-ph",
    doi = "10.1140/epjc/s10052-025-15280-w",
    journal = "Eur. Phys. J. C",
    volume = "86",
    number = "2",
    pages = "100",
    year = "2026"
}

@article{Ozdem:2022iqk,
    author = {{\"O}zdem, Ula{\c{s}}},
    title = "{Magnetic moments of pentaquark states in light-cone sum rules}",
    doi = "10.1140/epja/s10050-022-00700-2",
    journal = "Eur. Phys. J. A",
    volume = "58",
    number = "3",
    pages = "46",
    year = "2022"
}

@article{Ozdem:2023htj,
    author = {{\"O}zdem, Ula{\c{s}}},
    title = "{Electromagnetic properties of D{\textasciimacron}({\textasteriskcentered}){\ensuremath{\Xi}}c', D{\textasciimacron}({\textasteriskcentered}){\ensuremath{\Lambda}}c, D{\textasciimacron}s({\textasteriskcentered}){\ensuremath{\Lambda}}c and D{\textasciimacron}s({\textasteriskcentered}){\ensuremath{\Xi}}c pentaquarks}",
    eprint = "2303.10649",
    archivePrefix = "arXiv",
    primaryClass = "hep-ph",
    doi = "10.1016/j.physletb.2023.138267",
    journal = "Phys. Lett. B",
    volume = "846",
    pages = "138267",
    year = "2023"
}

@article{Ozdem:2024suc,
    author = {{\"O}zdem, Ula{\c{s}}},
    title = "{Insight into the nature of the $P_{c}(4457)$ and related pentaquarks}",
    eprint = "2409.09449",
    archivePrefix = "arXiv",
    primaryClass = "hep-ph",
    doi = "10.1140/epjc/s10052-025-14323-6",
    journal = "Eur. Phys. J. C",
    volume = "85",
    number = "6",
    pages = "624",
    year = "2025"
}

@article{Mutuk:2024ach,
    author = "Mutuk, Halil",
    title = "{Magnetic moments of hidden-charm pentaquarks in the diquark{\textendash}diquark{\textendash}antiquark scheme}",
    eprint = "2411.16486",
    archivePrefix = "arXiv",
    primaryClass = "hep-ph",
    doi = "10.1016/j.cjph.2025.07.030",
    journal = "Chin. J. Phys.",
    volume = "97",
    pages = "1406--1414",
    year = "2025"
}

@article{Marse-Valera:2022khy,
    author = "Mars{\'e}-Valera, J. A. and Magas, V. K. and Ramos, A.",
    title = "{Double-Strangeness Molecular-Type Pentaquarks from Coupled-Channel Dynamics}",
    eprint = "2210.02792",
    archivePrefix = "arXiv",
    primaryClass = "hep-ph",
    doi = "10.1103/PhysRevLett.130.091903",
    journal = "Phys. Rev. Lett.",
    volume = "130",
    number = "9",
    pages = "091903",
    year = "2023"
}

@article{Marse-Valera:2024apc,
    author = "Mars{\'e}-Valera, J. A. and Magas, V. K. and Ramos, A.",
    title = "{Double strangeness pentaquarks and other exotic hadrons in the {\ensuremath{\Xi}}b{\textrightarrow}{\ensuremath{\Xi}}J/{\ensuremath{\Psi}}{\ensuremath{\phi}} decay}",
    eprint = "2410.10682",
    archivePrefix = "arXiv",
    primaryClass = "hep-ph",
    doi = "10.1103/PhysRevD.111.054020",
    journal = "Phys. Rev. D",
    volume = "111",
    number = "5",
    pages = "054020",
    year = "2025"
}

@article{Ortega:2022uyu,
    author = "Ortega, Pablo G. and Entem, David R. and Fernandez, Francisco",
    title = "{Strange hidden-charm P{\ensuremath{\psi}}s{\ensuremath{\Lambda}}(4459) and P{\ensuremath{\psi}}s{\ensuremath{\Lambda}}(4338) pentaquarks and additional P{\ensuremath{\psi}}s{\ensuremath{\Lambda}}, P{\ensuremath{\psi}}s{\ensuremath{\Sigma}} and P{\ensuremath{\psi}}ssN candidates in a quark model approach}",
    eprint = "2210.04465",
    archivePrefix = "arXiv",
    primaryClass = "hep-ph",
    doi = "10.1016/j.physletb.2023.137747",
    journal = "Phys. Lett. B",
    volume = "838",
    pages = "137747",
    year = "2023"
}

@article{Wang:2025pjt,
    author = "Wang, Zhi-Gang and Liu, Yang",
    title = "{Analysis of the hidden-charm pentaquark candidates in the $J/ψΞ$ mass spectrum via the QCD sum rules}",
    eprint = "2511.13067",
    archivePrefix = "arXiv",
    primaryClass = "hep-ph",
    doi = "10.1016/j.nuclphysb.2026.117456",
    journal = "Nucl. Phys. B",
    volume = "1027",
    pages = "117456",
    year = "2026"
}

@article{CMS:2024vnm,
    author = "Hayrapetyan, Aram and others",
    collaboration = "CMS",
    title = "{Observation of the $\Lambda_\text{b}^0\to J/\psi\Xi^-K^+$ decay}",
    eprint = "2401.16303",
    archivePrefix = "arXiv",
    primaryClass = "hep-ex",
    reportNumber = "CMS-BPH-22-002, CERN-EP-2024-006",
    doi = "10.1140/epjc/s10052-024-13114-9",
    journal = "Eur. Phys. J. C",
    volume = "84",
    number = "10",
    pages = "1062",
    year = "2024"
}

@article{LHCb:2025lhk,
    author = "Aaij, R. and others",
    collaboration = "LHCb",
    title = "{Observation of the ${{{\varLambda } ^0_{b}} \!\rightarrow {{J \hspace{-1.66656pt}/\hspace{-1.111pt}\psi }} {{\varXi } ^-} {{K} ^+} }$ and ${{{{\varXi } ^0_{b}} \!\rightarrow {{J \hspace{-1.66656pt}/\hspace{-1.111pt}\psi }} {{\varXi } ^-} {{\pi } ^+} }}$ decays}",
    eprint = "2501.12779",
    archivePrefix = "arXiv",
    primaryClass = "hep-ex",
    reportNumber = "CERN-EP-2024-337 LHCb-PAPER-2024-049, CERN-EP-2024-337, LHCb-PAPER-2024-049",
    doi = "10.1140/epjc/s10052-025-14129-6",
    journal = "Eur. Phys. J. C",
    volume = "85",
    number = "7",
    pages = "812",
    year = "2025"
}

@book{Ali:2019roi,
    author = "Ali, Ahmed and Maiani, Luciano and Polosa, Antonio D.",
    title = "{Multiquark Hadrons}",
    doi = "10.1017/9781316761465",
    isbn = "978-1-316-76146-5, 978-1-107-17158-9, 978-1-316-77419-9",
    publisher = "Cambridge University Press",
    month = "6",
    year = "2019"
}

\end{document}